\documentclass[onefignum,onetabnum,11pt]{siamart220329}

\usepackage{amsmath,amssymb,amsfonts,color}
\usepackage{bm}
\usepackage[margin=3cm]{geometry}

\usepackage{amsfonts}
\usepackage{graphicx}
\usepackage{epstopdf}
\usepackage{algorithmic}
\ifpdf
  \DeclareGraphicsExtensions{.pdf,.png,.jpg}
\else
  \DeclareGraphicsExtensions{.eps}
\fi

\usepackage{subfig}

\newsiamremark{remark}{Remark}
\newsiamremark{hypothesis}{Hypothesis}
\crefname{hypothesis}{Hypothesis}{Hypotheses}
\newsiamthm{claim}{Claim}

\headers{Non-local models and convolution}{T. Nagel, T. Gerasimov, and D. Kern}

\title{Neighbourhood watch in mechanics: non-local models and convolution\thanks{
This article is currently under peer review.
\funding{This work was funded by the German Science Foundation (DFG) under grants NA1528/2-1 and MA4450/5-1.}}} 

\author{Thomas Nagel\thanks{Geotechnical Institute, TU Bergakademie Freiberg 
  (\email{thomas.nagel@ifgt.tu-freiberg.de}, \url{https://tu-freiberg.de/en/soilmechanics})}, 
\and Tymofiy Gerasimov\thanks{Department of Environmental Informatics, Helmholtz Centre for Environmental Research -- UFZ},   
\and Dominik Kern\thanks{Geotechnical Institute, TU Bergakademie Freiberg}
}

\usepackage{amsopn}

\usepackage{rotating}
\usepackage{tikz}
\usepackage{pgfplots}
\pgfplotsset{compat=1.17} 
\usetikzlibrary{math}
\usepackage{siunitx}
\usepgfplotslibrary{groupplots,dateplot}
\newlength\figH
\newlength\figW
\tikzset{
DKspring/.pic={
\coordinate (half_up) at (0.5*0.125*#1-0.5*0.125*2, 0.5*0.125*10-0.5*0.125*#1); 
\coordinate (full_up)   at ( 0.125*#1-    0.125*2,     0.125*10-    0.125*#1);
\coordinate (full_down) at ( 0.125*#1-    0.125*2,    -0.125*10+    0.125*#1);
\draw (0, 0) -- ++(1, 0) -- ++(half_up)
    -- ++(full_down) -- ++(full_up) 
    -- ++(full_down) -- ++(full_up)
    -- ++(full_down) -- ++(full_up)
    -- ++(full_down) -- ++(half_up)
    -- ++(1, 0);
    },   
DKdashpot/.pic={
\coordinate (upper_end) at (#1-0.5, 0.5);
\coordinate (lower_end) at (#1-0.5,-0.5);
\coordinate (upper_pos) at (#1-1, 0.5);
\coordinate (lower_pos) at (#1-1,-0.5);
\coordinate (center_pos) at (#1-1, 0.0);
\coordinate (center_end) at (#1, 0.0);
\draw (0, 0) -- ++(1, 0);
\draw (upper_end) -- (1, 0.5) -- (1, -0.5) -- (lower_end);
\draw (center_pos) -- (center_end);
\draw (upper_pos) -- (lower_pos);
    }
}

\ifpdf
\hypersetup{
  pdftitle={Neighbourhood watch in mechanics: non-local models and convolution},
  pdfauthor={Thomas Nagel, Tymofiy Gerasimov, Dominik Kern}
}
\fi

\makeatletter
\newcommand\footnoteref[1]{\protected@xdef\@thefnmark{\ref{#1}}\@footnotemark}
\makeatother

\begin{document}

\maketitle

\begin{abstract}
This paper is intended to serve as a low-hurdle introduction to non-locality for graduate students and researchers with an engineering mechanics or physics background who did not have a formal introduction to the underlying mathematical basis. We depart from simple examples motivated by structural mechanics to form a physical intuition and demonstrate non-locality using concepts familiar to most engineers. We then show how concepts of non-locality are at the core of one of the moste active current research fields in applied mechanics, namely in phase-field modelling of fracture. From a mathematical perspective, these developments rest on the concept of convolution both in its discrete and in its continuous form. The previous mechanical examples may thus serve as an intuitive explanation of what convolution implies from a physical perspective. In the supplementary material we highlight a broader range of applications of the concepts of non-locality and convolution in other branches of science and engineering by generalizing from the examples explained in detail in the main body of the article.
\end{abstract}

\begin{keywords}
convolution, linear differential equations, non-local theories, damage mechanics
\end{keywords}





\section{Introduction}
Advanced engineering theories always rest on mathematical concepts which---unless a bridge to knowledge acquired previously can be established---may look obscure, non-obvious or non-intuitive to a non-mathematician. Conversely, it may not be easy for mathematicians to convey useful (though, in many cases, abstract) concepts in simple words, without their usual ``jargon''. Luckily for both communities (engineers and applied mathematicians), it is almost always possible to find a simple example from engineering practice that would provide the most intriguing insights into mathematical concepts---the converse is, of course, also true. This is expressed in a famous quote attributed to Leonardo da Vinci:
\vspace{0.3cm}
\begin{quote}
    ``Mechanics is the paradise of the mathematical sciences because by means of it one comes to the fruits of mathematics.''
\end{quote}
\vspace{0.3cm}
As mechanicians, we believe this is very true. Once an example based on an intuitive reasoning is at hand, it becomes easier to comprehend a new theory or concept. From there, the path to abstraction and to a wealth of other applications of the same approach opens up. In this article, we address non-locality as an important concept in engineering and convolution as the underlying mathematical tool opening up a vast realm of other applications.

But all beginnings are difficult. Consider a PhD student studying the failure of materials---one of the prime tasks of solid mechanics---who consequently starts hearing and getting familiar with notions of fracture, damage, localization and other related ideas. It does not take long before the student comes across the concept on a non-local constitutive law (cf. \autoref{sec:nonlocal_material}) which aims at describing so-called  process-zone effects of material degradation, and also serves as a means of regularization of an ill-posed numerical model\footnote{Non-local theories appear in other branches of science and engineering as well, e.g. diffusion theories, quantum mechanics, etc. Here, we remain in the realm of the mechanics of materials and structures. Other applications are briefly discussed in the supplementary material which we recommend for further insights into discrete and continuous convolution.}. 
But what does this mean, ``non-local''? Why should the constitutive behaviour at a material point suddenly depend on some larger material volume in the vicinity of this point in departure of what the student was taught hitherto? The student may discover some vague explanation of how large molecules, electrical fields, microfissures etc. interact on physical length scales that exceed locality in the sense of the interaction of a point only with its immediate neighbourhood. Some students may content themselves with this mental image, others may remain puzzled. Especially, because the choice of the variable to be ``made non-local'' appears to be driven by practical considerations from the perspective of numerical modelling, in many cases, rather than by physical necessity. Aside from the choice of variable, the issue of the kernel function in the non-local integration routines remains. What is its origin---is it purely mathematical or can it be enriched with a physical interpretation? Having at hand a simple and intuitive example of non-locality and convolution integrals would certainly be helpful for gaining confidence in these theories.

There is good reason to hope for such an accessible example: Many students will have come into contact with non-local effects as early as in their undergraduate degrees, although it is commonly not pointed out to them at the time and thus may have gone unnoticed. Taking the example of an elastically bedded plate as point of departure, we show how structural effects can lead to non-locality and the appearance of convolution integrals with kernel functions (\autoref{sec:boussinesq}). We then show how changing the kernel functions can lead to the appearance of both simpler (local) and more general (mixed) theories. 

In the second part of the article, insights are given into the use of non-locality in a currently quite active field of research in various branches of engineering mechanics: the use of phase-fields to represent fracture-mechanical concepts (\autoref{sec:nonlocal_material}). In this section we hope to link the didactic nature of the present article to research relevant to many graduate students and scientists, as well as to provide another view point, another angle and thus more insight into non-locality.

As pointed out several times by now, we make wide use of convolution integrals. For those whose curiosity was sparked by this concept we provide a supplement to illustrate the use of convolution integrals on a wider set of examples from various branches of mathematics, science and engineering to demonstrate, how the understanding provided by one simple example can help us generalize ideas to other problem classes, such as sums of random variables, 
linear time-invariant systems and visco-elasticity (\autoref{sec_convolution}).

\section{A basic example from engineering mechanics}
\label{sec:boussinesq}

Simply supported beams and plates, that is, thin, one- and two-dimensional structures carrying out-of-plane loads predominantly by bending modes, are at the base of most engineering mechanics courses. More intricate structures such as, e.g., elastic foundations intensively studied in geotechnical engineering applications can be represented by means of these basic structural models. It can be shown that the way a foundation deforms as well as the internal reactions (e.g. bending moments) it is subjected to depend, in fact, on the interaction between the foundation and the supporting soil.

In the mechanical context, this is a natural example of a so-called non-local behaviour, as will be shown in the sequel. From the mathematical point of view such an interaction is described via the notion of convolution (rather, convolution integral) also making use of a so-called kernel function, the precise meaning of which will be given below.

In the following, we depart from Boussinesq's fundamental solution of a point force acting on an elastic half space. This solution can be generalised to surface tractions by integration exploiting the powerful superposition principle, allowing the study of the interaction of an elastic plate with an elastic half space. This naturally leads to a non-local model with an intuitive explanation. Thereby, we also introduce the notions of kernel functions and convolution integrals by generalizing from the familiar principle of superposition. Finally, we demonstrate how the choice of kernel functions can recover local theories, e.g. a plate bedded on a Winkler-type half space, the probably simplest approximation of an elastic half-space.

\subsection{A point load applied to elastic half space}
\label{sec_point_halfspace}

Let us consider a 3-dimen- sional half-space such that $(x,y,z)\in\mathbb{R}\times\mathbb{R}\times\mathbb{R}^{+}_0$. 
We assume its mechanical properties are described by an isotropic elastic medium with the Young modulus $E$ and the Poisson ratio $\nu$. For the sake of convenience, let the positive direction of $z$ point downwards. Let now $F_z$ be a vertical force applied at point $(0,0,0)$ in the positive direction of $z$ (compressive load on the surface). The stress distribution and the displacements of the half-space under the point load are described by what is called Boussinesq's solution \cite{boussinesq1885application}; see the appendix of Verruijt's textbook \cite{verruijt2017introduction} for an instructive solution using potentials.

According to this solution, the increments of stresses (compression positive according to geomechanical sign convention) which occur due to the applied load $F_z$ read as follows:
\begin{subequations}
\label{eq:sigma_boussinesq}
\begin{align}
    \Delta \sigma_{xx} &= \frac{3F_z}{2\pi R^2} \left[ \frac{x^2 z}{R^3} - \frac{1-2\nu}{3} \left( \frac{(x^2 - y^2) R}{r^2 (R + z)} + \frac{y^2 z}{R r^2}\right) \right],
    \\
    \Delta \sigma_{yy} &= \frac{3F_z}{2\pi R^2} \left[ \frac{y^2 z}{R^3} - \frac{1-2\nu}{3} \left( \frac{(y^2 - x^2) R}{r^2 (R + z)} + \frac{x^2 z}{R r^2}\right) \right],
    \\
    \Delta \sigma_{zz} &= \frac{3F_z}{2\pi R^2} \frac{z^3}{R^3},
    \\
    \Delta \sigma_{xy} &= \frac{3F_z}{2\pi R^2} \left[ \frac{x y z}{R^3} - \frac{1-2\nu}{3} \frac{xy(2R+z)}{R(R+z)^2} \right],
    \\
    \Delta \sigma_{yz} &= \frac{3F_z}{2\pi R^2} \frac{yz^2}{R^3},
    \\
    \Delta \sigma_{xz} &= \frac{3F_z}{2\pi R^2} \frac{xz^2}{R^3},
\end{align}
\end{subequations}
where $r: = \sqrt{x^2 + y^2}$  and $R: = \sqrt{x^2 + y^2 + z^2}$.

\begin{figure}[htb]
	\subfloat[Isobars of vertical stress increment $\Delta \sigma_{zz}$ beneath the point load.]{\includegraphics[width=0.45\textwidth]{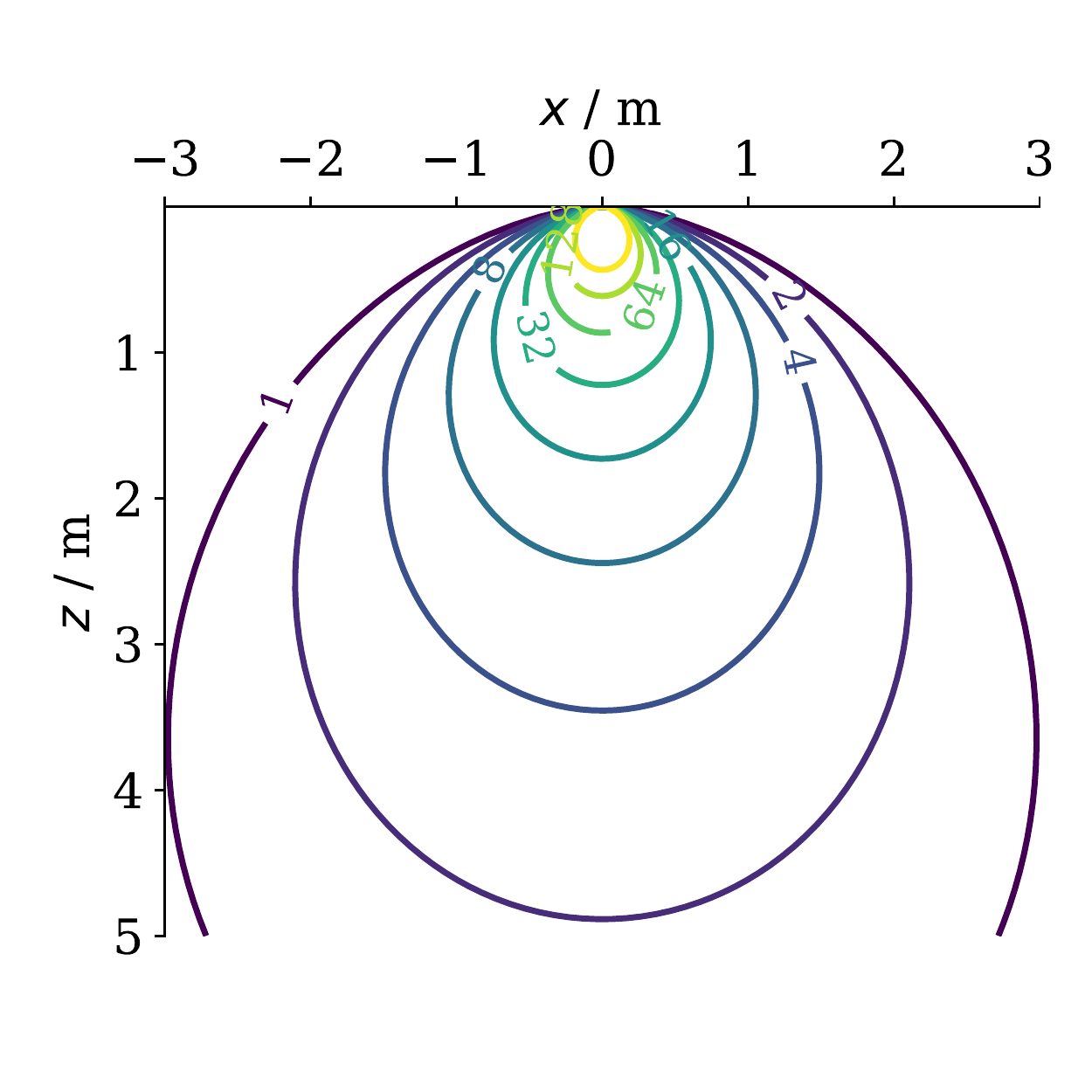} 
	\label{fig:Boussinesq_isobars}}
	\hfill
	\subfloat[Isobars and principal directions of largest compressive principal stress in heavy elastic half space: $\sigma_{ij} = \sigma_{0,ij} + \Delta \sigma_{ij}$.]{\includegraphics[width=0.49\textwidth]{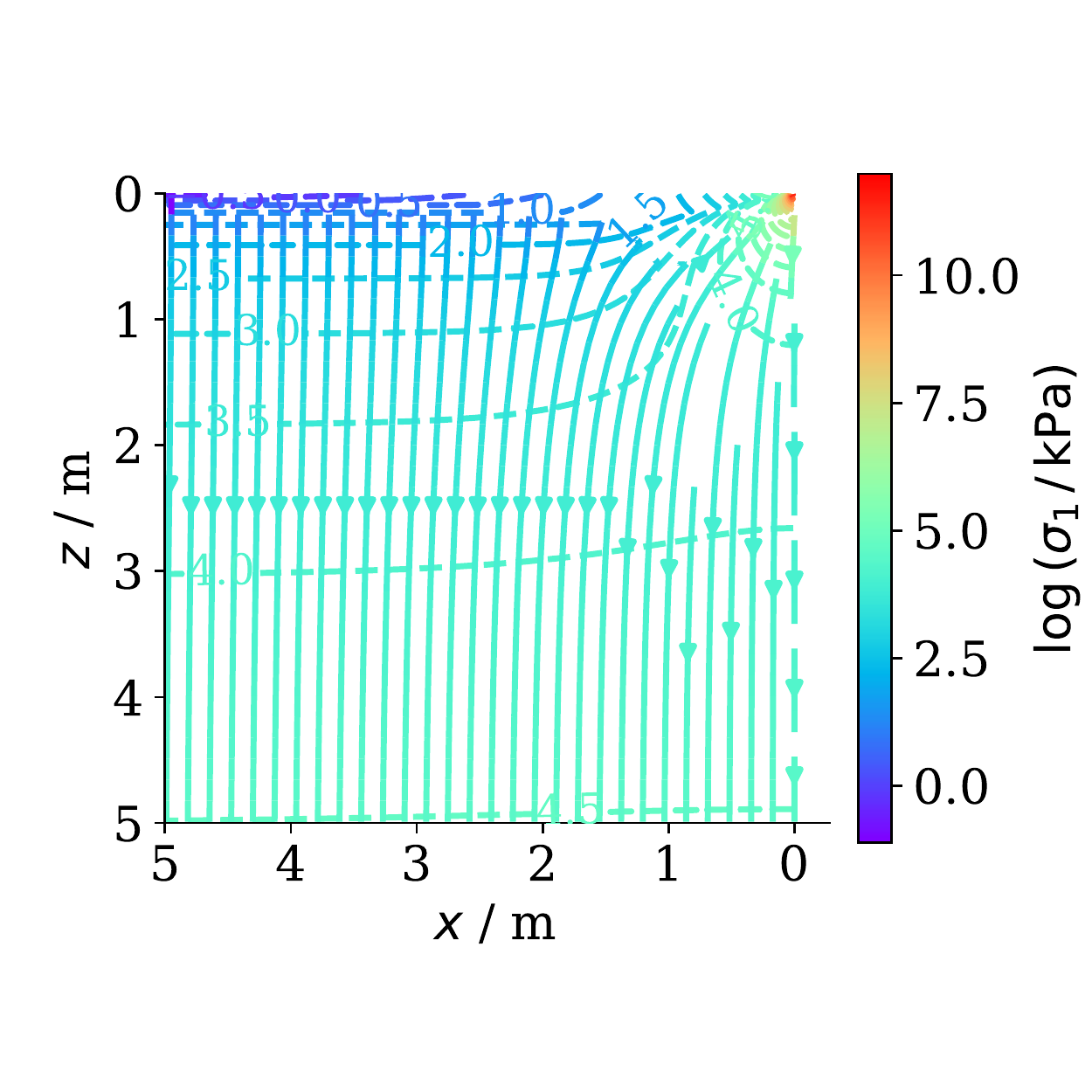} 
	\label{fig:Boussinesq_total}}
	\caption{Stresses in elastic half space subjected to vertical point load $F_z$. Stress increments from \autoref{eq:sigma_boussinesq} (a) and total stresses due to both load and gravitational in-situ stress (b).}
	\label{fig:Boussinesq}
\end{figure}

The solution is illustrated in \autoref{fig:Boussinesq}. For the total stress plot we assume a gravitational initial stress field of the form $\sigma_{0,ij} = \text{diag}\,[\nu/(1-\nu),\nu/(1-\nu),1] \gamma z$, where $\gamma$ is the specific weight of the half space, so that $\sigma_{ij} = \sigma_{0,ij} + \Delta \sigma_{ij}$.
The superposition principle holds for linear systems and allows us to add the responses (stresses) for individual loads to obtain the total solution.
Keep in mind that linearity is nothing to be taken for granted, as Steven Strogatz points out felicitously: ``listening to two of your favorite songs at the same time does not double the pleasure'' \cite{strogatz2018nonlinear}.

In the following, we are going to make full use of this principle when it comes to the action of several point loads and their generalization to distributed line or surface loads.

Once the stresses are calculated, we can compute strains using the linear elasticity assumption and, as a result, end up with the corresponding displacements by integration. The $u_z$-component of the displacement vector at the top surface is of particular interest for geotechnical engineers and is called the settlement. It reads

\begin{equation}
    s(x,y) = u_z(x,y)|_{z=0} = \int \limits_{z=0}^{z=\infty} \epsilon_{zz} \text{d}z,
    \label{eq:settle}
\end{equation}
with $\epsilon_{zz}$ given by\footnote{We neglect the influence of groundwater here such that total and effective stresses coincide.} 

\begin{equation}
    \epsilon_{zz} = \frac{1}{E} \left[ \Delta \sigma_{zz} - \nu (\Delta \sigma_{xx} + \Delta \sigma_{yy})  \right].
    \label{eq:epszz}
\end{equation}

Substituting \autoref{eq:sigma_boussinesq} into \autoref{eq:epszz} and using \autoref{eq:settle}, we arrive at the expression for the settlement of the elastic half-space,  one of the Boussinesq solutions:

\begin{equation}
    s(r) = u_z(r,z=0) = \frac{F_z(1-\nu^2)}{\pi E r}.
    \label{eq:trough}
\end{equation}

Note that the constructed $s$ is singular (blows up, is infinite) at $r=0$ as a consequence of the point load idealization and the limitation to linear elasticity. However, since the solution was derived under the assumptions of linear kinematics and linear isotropic elasticity, the superposition principle is at our disposal to obtain stress distributions for more complex surface loads. This will be demonstrated in the next section.

\subsubsection{Discrete superposition}

\begin{figure}[htb]
	\subfloat[Superposition of 3 point loads]{\includegraphics[width=0.491\textwidth]{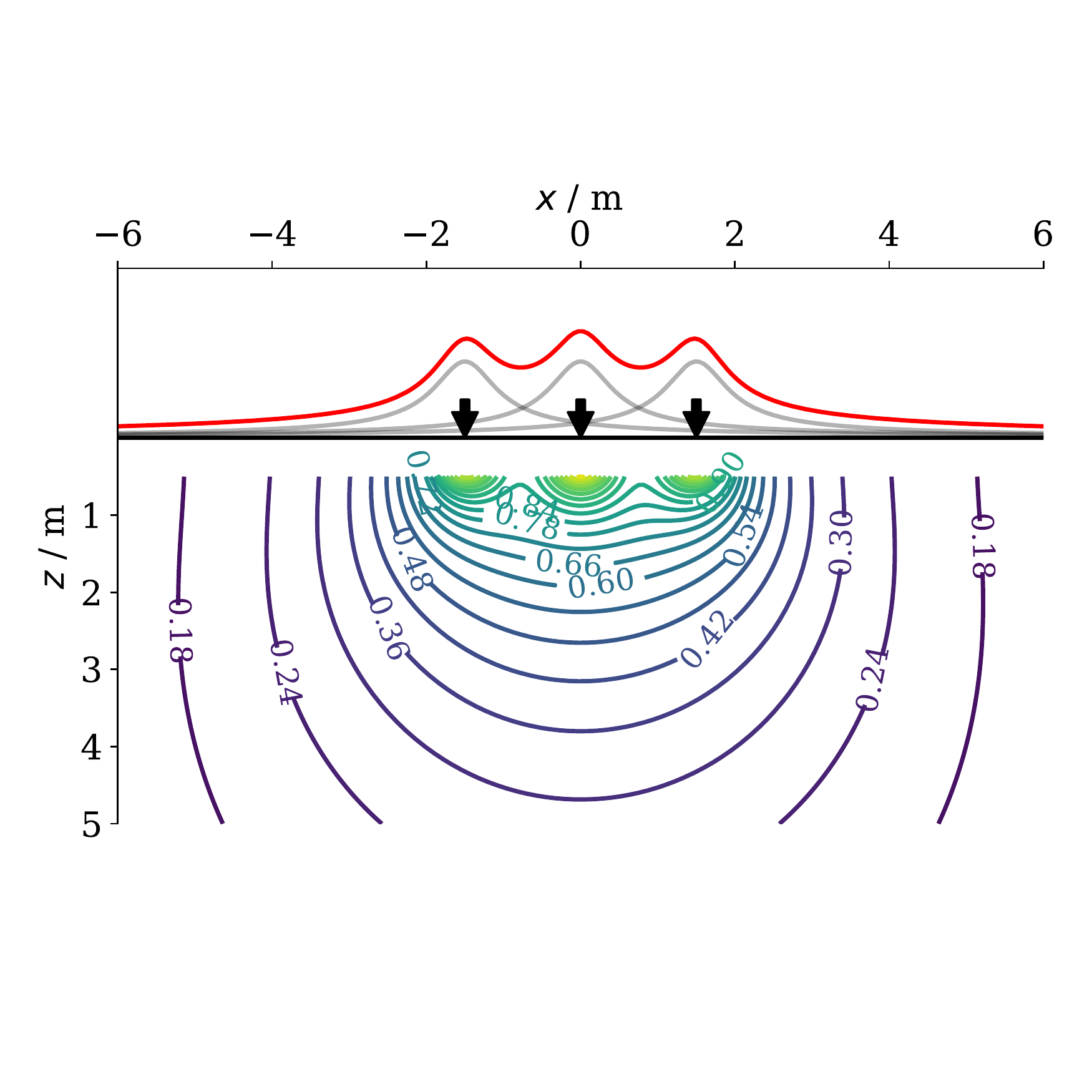} 
	\label{fig:superposition_3}}
	\hfill
	\subfloat[Superposition of 10 point loads]{\includegraphics[width=0.491\textwidth]{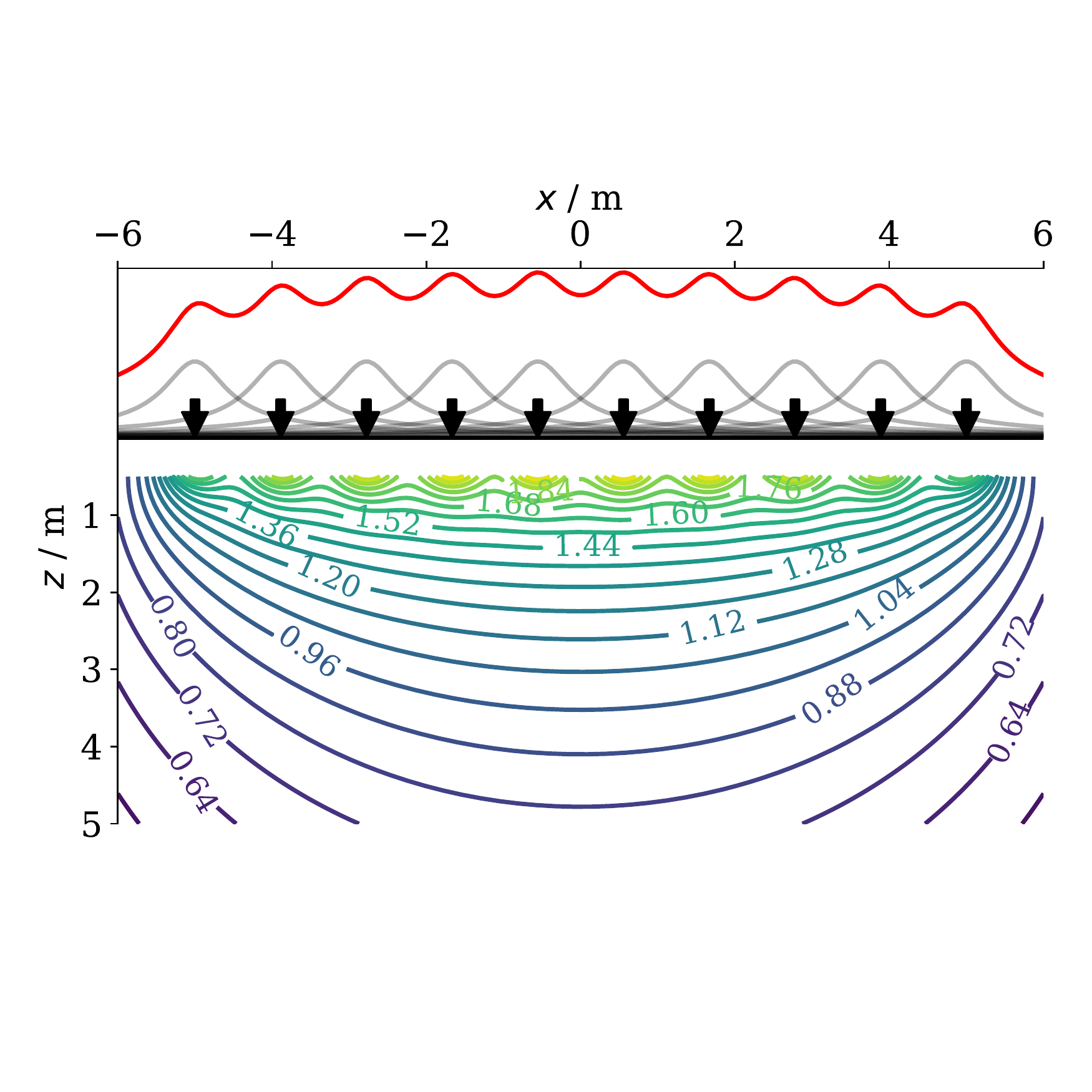} 
	\label{fig:superposition_10}}
	\caption{Superposition of point loads. The bottom part of each figure shows vertical stress isobars for $z>0.5$\,m. The top part shows settlement distributions of each point load at $z=0.5$\,m (gray) and the resulting superposition (red). As the number of loads increases, a type of solution characteristic for a distributed line load is approached.}
	\label{fig:Superposition_loads}
\end{figure}

The superposition of \autoref{eq:sigma_boussinesq} and \autoref{eq:settle} for sets of point loads acting on the surface of the elastic half space proceeds by simple summation and by shifting the solutions according to the point of attack of each individual force. 

\begin{equation}
    s(x,y) = \sum \limits_{i=1}^n s(x - x_\text{i},y - y_\text{i}) = \frac{F_z(x_\text{i},y_\text{i}) (1-\nu^2)}{\pi E \sqrt{(x - x_\text{i})^2 + (y - y_\text{i})^2}}
\end{equation}

The results in Fig.~\ref{fig:Superposition_loads} have been generated with an interactive Jupyter notebook\footnote{\label{jupyter}Available here:\\  \url{https://mybinder.org/v2/gh/nagelt/Teaching_Scripts/SIAM?labpath=superposition_Boussinesq.ipynb}}, which can be run interactively on \url{mybinder.org}, a tool very useful for facilitating understanding \cite{Kern2022}. 

\subsubsection{Continuous superposition}

Transitioning from the superposition of discrete loads to continuous loads is achieved by integrating distributed line-loads in one horizontal direction instead of summing up individual point loads. In fact, we transition to infinitesimal point loads juxtaposed ever so closely. Integrating in the second horizontal direction by transitioning to surface tractions then yields the stress distribution under rectangular pressure loads. This solution as well as the loading by stiff plates can likewise be viewed in the above mentioned Juypter notebook\footnoteref{jupyter}. To limit redundancy, it is given and used in the following section, cf. \autoref{eq:settlement_increment} and \autoref{eq:s_nonlocal}.

\subsection{Elastic plate on elastic half space---arriving at non-locality}
For a basic understanding of the elastically bedded plate theory, we depart from the differential equation for a thin plate, which shows an obvious analogy to beam theory. In fact, Kirchhoff's plate theory can be seen as a direct extension of Bernoulli's beam theory to plates. For homogeneous, isotropic, thin plates bending under transversal loading we depart from the following bi-potential equation

\begin{equation}
    \Delta \Delta w = \frac{\partial^4 w}{\partial x^4} + 2\frac{\partial^4 w}{\partial x^2 \partial y^2} + \frac{\partial^4 w}{\partial y^4} = \frac{q(x,y) - \sigma_0(x,y)}{K}
    \label{eq:plate}
\end{equation}

with the plate stiffness $K = {E_\text{plate} h^3}/{[12(1-\nu_\text{plate}^2)]}$, the plate thickness $h$ and the elastic constants of the plate, $E_\text{plate}$ and $\nu_\text{plate}$.  The known external normal traction (load) on the plate is described by $q(x,y)$, while $\sigma_0(x,y)$ represents the yet unknown soil reaction pressure.

We assume continuous contact between foundation and soil, in other words, soil settlement $s(x,y)$ and the bending profile $w(x,y)$ of the plate must be equal (no penetration, no lift-off)

\begin{equation}
    w(x,y) = s(x,y)
    \label{eq:ws}
\end{equation}

So far we've looked at the elastic behaviour of the plate loaded from above by $q(x,y)$ and from below by $\sigma_0(x,y)$ only, as described by \autoref{eq:plate}. Since $\sigma_0(x,y)$ is not yet known, we make use of the constraint \autoref{eq:ws} and obtain additional information on the link between $s(x,y)$ and $\sigma_0(x,y)$ by considering the deformation of the elastic half space, the soil. 

\begin{figure}
    \centering
    \includegraphics[width=0.6\textwidth]{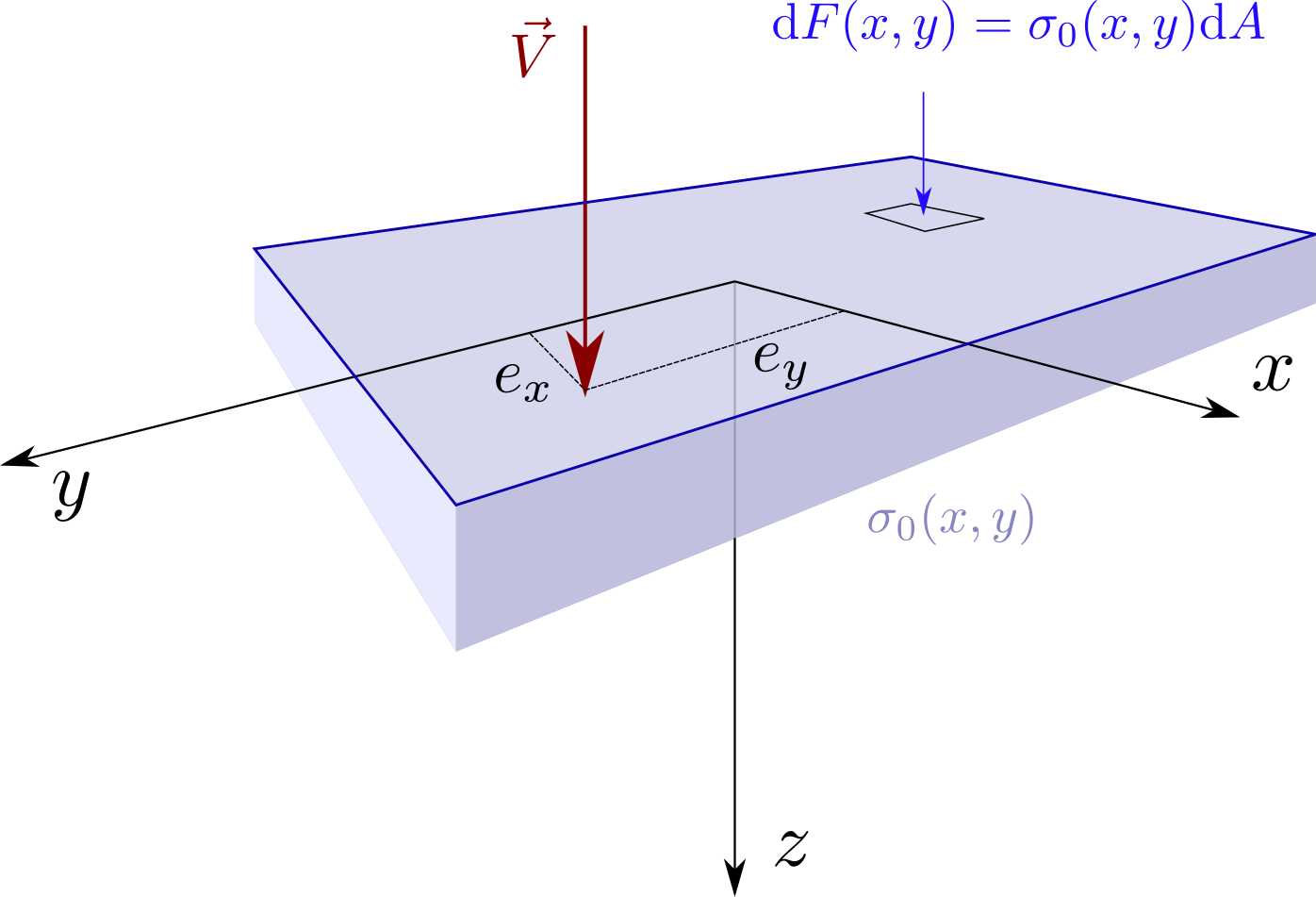}
    \caption{Illustration of the normal traction $\sigma_0(x,y)$ acting in the contact zone between plate bottom and soil surface. This traction is in equilibrium with corresponding reaction forces. At each point $(x,y)$, infinitesimal forces $\text{d}F$ result on infinitesimal area elements $\text{d}A$. $\vec{V}$ indicates a statically equivalent reaction force (e.g. the loading of the foundation) which may be eccentric.}
    \label{fig:Gleichgewicht_Sohlfuge}
\end{figure}

The key is to consider the effect of the traction field $\sigma_0(x,y)$ as a superposition of infinitesimal neighbouring forces $\text{d}F_z$ as indicated in \autoref{fig:Gleichgewicht_Sohlfuge}. In other words, we consider the effect of the soil pressure $\sigma_0$ on an infinitesimal surface element $\text{d}x_\text{m} \times \text{d}y_\text{m}$ at the location $(x_\text{m},\,y_\text{m})$. The resulting infinitesimal load $\text{d}F_z(x_\text{m},y_\text{m}) = \sigma_0(x_\text{m},y_\text{m}) \text{d}x_\text{m} \text{d}y_\text{m}$, according to Boussinesq, calls for a settlement trough centred around the force application point of the form (compare \autoref{eq:trough})

\begin{equation}
\begin{split}
    \text{d} s(x - x_\text{m},y - y_\text{m}) &= \frac{\text{d}F_z(x_\text{m},y_\text{m}) (1-\nu^2)}{\pi E \sqrt{(x - x_\text{m})^2 + (y - y_\text{m})^2}}
    \\
    &= \frac{\sigma_0(x_\text{m},y_\text{m}) (1-\nu^2)}{\pi E \sqrt{(x - x_\text{m})^2 + (y - y_\text{m})^2}} \text{d}x_\text{m}\text{d}y_\text{m}
\end{split}
\label{eq:settlement_increment}
\end{equation}

The natural curvature of the settlement trough due to this point load differs from the corresponding natural curvature of the plate determined its elastic length, i.e. its elastic properties in bending determined by the plate stiffness $K$. In other words, to maintain the constraint \autoref{eq:ws}, additional interaction tractions must ensure $w(x,y)=s(x,y)$ at every point of the plate. Sacrificing rigor for the sake of the image we can say that the infinitesimal force acting locally at point $(x_\text{m},\,y_\text{m})$ pushes the soil away from the plate also at neighbouring points reducing contact pressures in its vicinity. It becomes clear, that all infinitesimal point loads must somehow interact due to this \textit{non-local} effect. The strength of this interaction between two point loads should intuitively decay with the distance between between them.

Let's formalize this. The settlement troughs of all neighbouring infinitesimal point loads overlap to form the overall settlement trough:

\begin{equation}
    s(x,y) = \frac{1-\nu^2}{\pi E} \int \limits_{y_\text{m}} \int \limits_{x_\text{m}} \frac{\sigma_0(x_\text{m},y_\text{m})}{\sqrt{(x - x_\text{m})^2 + (y - y_\text{m})^2}} \text{d}x_\text{m}\text{d}y_\text{m}
    \label{eq:s_nonlocal}
\end{equation}

Here, the non-local effect becomes apparent. It is caused by the interaction of neighbouring soil elements that lead to loads distributing along stress trajectories in the subsurface, cf. \autoref{fig:Boussinesq}. This influence is formally represented in Eq.~\autoref{eq:s_nonlocal} via the Kernel function, or influence function, 

\begin{equation}
    f_\text{nl}(x,y;x_\text{m},y_\text{m}) = \frac{1}{\sqrt{(x - x_\text{m})^2 + (y - y_\text{m})^2}} 
    \label{eq:fnl}
\end{equation}

which captures the decay away from ($x_\text{m},\,y_\text{m}$) intuited earlier. With this, we can write

\begin{equation}
    s(x,y) = \frac{1-\nu^2}{\pi E} \int \limits_{y_\text{m}} \int \limits_{x_\text{m}} {\sigma_0(x_\text{m},y_\text{m})} f_\text{nl}(x,y;x_\text{m},y_\text{m}) \text{d}x_\text{m}\text{d}y_\text{m}
\end{equation}

The forefactor is recognized from \autoref{eq:trough}, and the influence function takes the role of the term $r^{-1}$ in \autoref{eq:trough} for the set of infinitesimal forces constituting the traction field.

This makes it clear that the contribution of the infinitesimal point load $\text{d}F_z$ applied at $(x_\text{m},\,y_\text{m})$ to the settlement at the point $(x,\,y)$ decreases with the distance between both points. In other words, the settlement at a point is dominated by the loads in the immediate vicinity of the point, but less so by points farther away. A lowering of the elastic slab at a certain point influences the boundary conditions acting on the slab in a certain environment around this point. Such an effect can be called non-local.\\

\textbf{Remark 1:} If one assumes a constant surface load intensity $q(x,y)=q_0$ and a negligible bending stiffness $K=0$ of the plate, then only the influence function $f$ remains in the integrand. Integration over a rectangular surface leads to the well-known solution of the settlement under a uniformly distributed rectangular load directly acting on a soil and is given in the Jupyter notebook linked in \autoref{jupyter} on page \pageref{jupyter}. This approach also opens up solution schemes based on piecewise constant $\sigma_0$-distributions used for tabulated solutions.\\

\textbf{Remark 2:} We arrive at the elastically bedded beam of length $l$ and width $b$ with the assumption that $\sigma_0$ is variable only in the $x$ direction. Thus we find another often used relation

\begin{equation*}
    s(x) = \frac{1-\nu^2}{\pi E} \int \limits_{-l/2}^{l/2} \sigma_0(x_\text{m}) f_\text{nl;b}(x;x_\text{m}) \text{d}x_\text{m}
\end{equation*}

with the kernel function 

\begin{equation*}
    f_\text{nl;b}(x;x_\text{m}) = \int \limits_{-b/2}^{b/2} \frac{\text{d}y}{\sqrt{(x - x_\text{m})^2 + y^2}}
\end{equation*}

\subsubsection{Changing the kernel function}
The kernel function above was not simply postulated. It was derived as a consequence of certain assumptions on the behaviour of the physical system. Nevertheless, we may ask what happens if we change the Kernel function in order to change the extend of non-locality or the manner in which neighbouring points interact. One may, for example, choose a strictly local interaction model by selecting Dirac delta distribution 

\begin{equation}
    f_\text{l}(x,y;x_\text{m},y_\text{m}) = \delta(x-x_\text{m})\delta(y-y_\text{m})
    \label{eq:fl}
\end{equation}

arriving at

\begin{equation}
    s(x,y) = \frac{1-\nu^2}{\pi E} \int \limits_{y_\text{m}} \int \limits_{x_\text{m}} \sigma_0(x_\text{m},y_\text{m}) \delta(x-x_\text{m})\delta(y-y_\text{m}) \text{d}x_\text{m}\text{d}y_\text{m} = \frac{1-\nu^2}{\pi E} \sigma_0(x,y)
\end{equation}
    
We observe a strictly local interaction, that is the settlement at any given point is a function of the load at this point without any influence of its vicinity. This corresponds to the subgrade reaction modulus method in geotechnical engineering and leads to a completely different model of stress distribution in the elastic half space. In fact, it corresponds to the introduction of a Winkler-type elastic half space \cite{winkler1867lehre}. In other words, the plate is now bedded on a set of mutually independent vertical springs.\\

\textbf{Remark 3:} To yield more realistic settlement shapes, some methods combine the two half space models from \autoref{eq:fnl} and \autoref{eq:fl} via their ``superposition'', that is

\begin{equation*}
    s(x,y) = \frac{1-\nu^2}{\pi E} \int \limits_{y_\text{m}} \int \limits_{x_\text{m}} \sigma_0(x_\text{m},y_\text{m}) \left[ \lambda f_\text{l} + (1-\lambda) f_\text{nl} \right] \text{d}x_\text{m}\text{d}y_\text{m}.
\end{equation*}
with $\lambda$ as the weighting factor, derived from an optimality condition.

This concludes the exemplary introduction of non-locality. We have seen that it naturally arises in engineering mechanics when studying the interaction of an elastic structual element with an elastic half space. Mathematically, \autoref{eq:s_nonlocal} is a convolution integral with the physically motivated kernel function \autoref{eq:fnl}. The concept of convolution is discussed more generally in the supplementary material. In the next section, we continue the theme of non-locality and provide some insights into its role in current research on the phase-field modelling of fracture mechanics.  

\section{Non-locality in damage and fracture mechanics}
\label{sec:nonlocal_material}

In this section, we want to present and illustrate two instances of the concept of {\em non-locality} realized mathematically via the concept of {\em convolution}, which are encountered in advanced mechanical topics like Continuum Damage Mechanics (CDM) and Linear Elastic Fracture Mechanics (LEFM). As the name suggests, both aim at describing and modeling material degradation and failure phenomena, but with  historically different origins, capabilities and numerical realizations. In the former case, development of the so-called {\em non-local} CDM formulations was driven to a large part by technical reasons, namely the need to regularize the ``unwanted'' localization effects inherent in the original {\em local} formulations which manifest themselves as non-physical numerical results obtained using finite element implementations \cite{Jirasek2007}. In case of LEFM, the relevant example of non-locality and convolution can be found specifically in the context of the variational approach to fracture, prevalent in the recent literature of phase-field modeling of fracture (PMF). Here, we will show that the corresponding governing equations of PMF, see \cite{Ambati2015review},  originally derived via variational principles can also be obtained departing from a convolution representation. Such an alternative derivation may justify a phenomenological approach for deriving the material (failure) models which are of a non-variational nature or, in other words, are variationally inconsistent.

\subsection{Non-local and gradient-enhanced formulations of Continuum Damage Mechanics}

The basic idea of classical continuum damage mechanics\footnote{The fundamentals of the CDM theory including an extensive overview can be found in the monographs of Kachanov \cite{Kachan} and Murakami \cite{Murak}.} is to model the loss of stiffness associated to mechanical degradation of (linearly elastic) materials by a scalar parameter $D$ according to the stress-strain relation $\bm\sigma(\bm u,D):=(1-D)\mathbb{C}:\bm\varepsilon$, where $D$ ranges from 0 (virgin material, with elastic stiffness $\mathbb{C}$) and 1 (completely damaged material, with no stiffness). To allow the damage variable $D$ to evolve, one must make it depend on some state variable which naturally evolves due to changing external loading applied to the mechanical system. Typically, such a state variable is, in turn, taken to be a function of the strain $\bm\varepsilon$.

\begin{figure}[!ht]
\begin{center}
\includegraphics[width=1.0\textwidth]{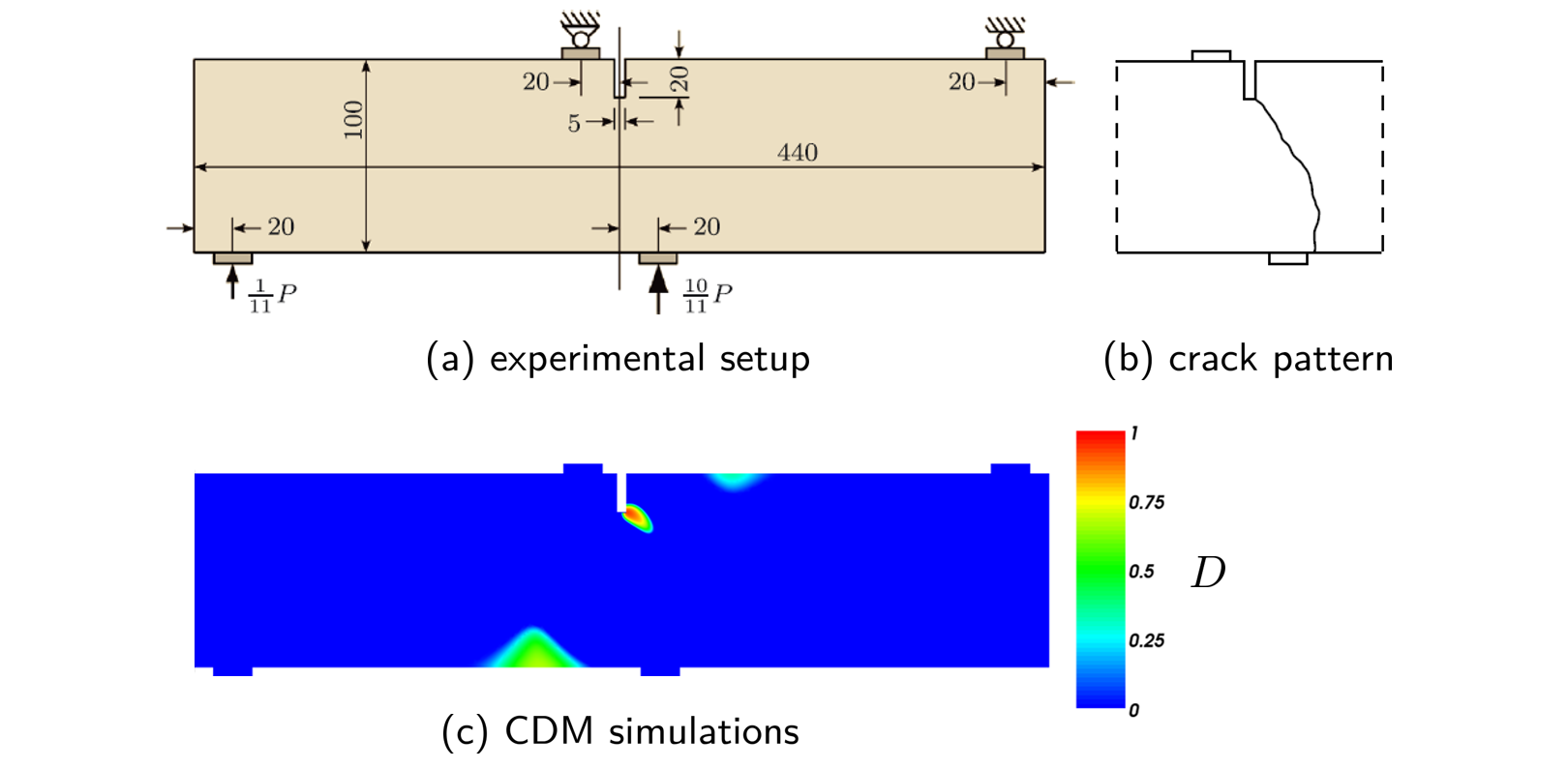}
\end{center}
\caption{(a) Experimental setup for the SEN concrete beam undergoing anti-symmetric four point bending \cite{Schlangen,Geers}, all dimensions in mm; (b) the experimentally observed failure (crack) pattern \cite{Schlangen}; (c) initiation and early stage of crack propagation using finite element CDM modeling. }
\label{fig:CDM}
\end{figure}

\autoref{fig:CDM} illustrates an application of the CDM to model crack initiation and propagation in a single edge notched (SEN) concrete beam subject to a particular type of four point bending loading. In the simulation results, three damaged zones within the specimen can be observed, but only one where the damage variable $D$ reaches 1 represents the zone of actual failure mimicking crack initiation and propagation.  

In the seminal CDM formulations termed {\em local}, the state variable is presented by the so-called equivalent strain $\widetilde{\varepsilon}=\widetilde{\varepsilon}(\bm\varepsilon)$ such that $D=D(\widetilde{\varepsilon})$. Multiple definitions of $\widetilde{\varepsilon}$ have been developed to properly account for different phenomena and drivers of damage initiation and progression the model is supposed to reproduce, see e.g.\ \cite{Peerl} for discussion. The two most commonly used models in engineering practice read
\begin{equation}
\widetilde{\varepsilon}:=\left\{
\begin{tabular}{l}
$\sqrt{\displaystyle\sum_{I=1}^3 \langle\varepsilon_I\rangle_{+}^2}$, \vspace{0.1cm} \\
$\displaystyle\frac{\kappa-1}{2\kappa(1-2\nu)}I_1(\bm\varepsilon)+\displaystyle\frac{1}{2\kappa}
\left[ \Big( \frac{\kappa-1}{1-2\nu}I_1(\bm\varepsilon) \Big)^2+ \displaystyle\frac{12\kappa}{(1+\nu)^2}J_2(\bm\varepsilon)^{} \right]^\frac{1}{2}$.
\end{tabular}
\right.
\label{LocEquivStr}
\end{equation}
proposed in \cite{Mazars} and \cite{deVree}, respectively. In \autoref{LocEquivStr}.a, $\langle\varepsilon_I\rangle_{+}:=\max(0,\varepsilon_I)$ stand for the positive part of the principal strains $\varepsilon_I$, $I=1,2,3$. In \autoref{LocEquivStr}.b, $\kappa\in\mathbb{R}$ is a dimensionless parameter, $I_1(\bm\varepsilon):=\mathrm{tr}(\bm\varepsilon)$ and $J_2(\bm\varepsilon):=\mathrm{tr}(\bm\varepsilon^2)-\frac{1}{3}\mathrm{tr}^2(\bm\varepsilon)$ are the first invariant of the strain tensor and the second invariant of the deviatoric strain tensor, respectively. Both definitions are specifically designed to enable the models to distinguish between tensile and compressive material damage, as well as to control model's sensitivity to certain deformation modes.

\begin{figure}[!ht]
\begin{center}
\includegraphics[width=1.0\textwidth]{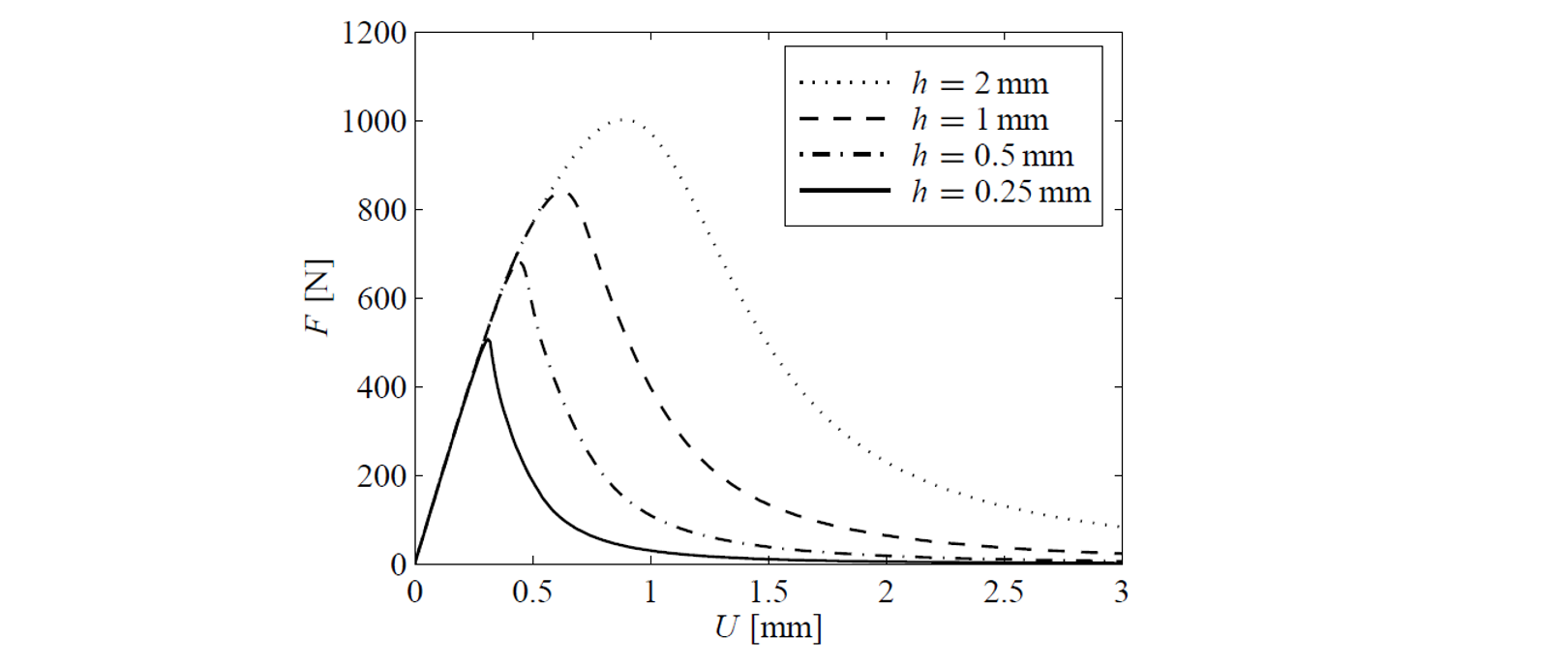}
\end{center}
\caption{Mesh-sensitivity of local CDM formulations: a significant change of the numerical results---here, the load-displacement curve $F$-$U$---is observed during refinement of the underlying mesh used in the computations, demonstrating no sign of usual convergence in the limit of $h\rightarrow 0$; the original figure is taken from \cite{Peerl}.}
\label{fig:CDM_ms}
\end{figure}

Unfortunately, even with physically justified equivalent strain definitions, local CDM formulations of the kind 
\begin{equation}
\left\{
\begin{tabular}{c}
$\bm\sigma(\bm u,D):= \bigl(1-D(\widetilde{\varepsilon})\bigr)\,\mathbb{C}:\bm\varepsilon$, \\[0.2cm]
$\widetilde{\varepsilon}$ given by \autoref{LocEquivStr} or similar,
\end{tabular}
\right.
\label{LocalCMD}
\end{equation}
are well-known to suffer from spurious mesh sensitivity (also termed mesh dependendency) when it comes to the finite element discretisation, thus exhibiting physically unrealistic simulations results. In particular, as stated in \cite{Peerl}:
\vspace{0.3cm}
\begin{quote}
 ``The growth of damage tends to localise in the smallest band that can be captured by the spatial discretisation. As a consequence, increasingly finer discretisation grids lead to crack initiation earlier in the loading history and to faster crack growth. In the limit of an infinite spatial resolution, the predicted damage band has a thickness zero and the crack growth becomes instantaneous. The response is then perfectly brittle, i.e., no work is needed to complete the fracture process. This nonphysical behaviour is caused by the fact that the localisation of damage in a vanishing volume is no longer consistent with the concept of a continuous damage field which forms the basis of the continuum damage approach.''
\end{quote}
\vspace{0.3cm}
\autoref{fig:CDM_ms}, which for the sake of consistency is taken from \cite{Peerl} as well, illustrates the situation in terms of the so-called load-displacement curve: one observes no sign of convergence for the simulation results in the usual finite element sense when  $h\rightarrow 0$. In simple words, the former finding is in contradiction with the fundamental finite element paradigm stating that ``the smaller the elements are, the more accuracy our numerical solution gains''. Here instead, with $h\rightarrow 0$, the physical predictions are really questionable, if not meaningless.

Mathematically, the undesired local model's behaviour occurs due to the fact that the damage parameter depends on the strain state defined only at individual points: indeed, the stipulation $\widetilde{\varepsilon}=\widetilde{\varepsilon}(\bm\varepsilon)$ with $\bm\varepsilon=\bm\varepsilon(\bm x)$ implies $\widetilde{\varepsilon}=\widetilde{\varepsilon}(\bm x)$. The natural idea to revise the situation is to replace $\widetilde{\varepsilon}$ in $D=D(\widetilde{\varepsilon})$ with a quantity which is no longer specified only at a point, but also ``feels'' and ``receives information'' from the neighboring ones. This can be, and actually is, realized using the concept of convolution alluded to earlier: one introduces the so-called non-local equivalent strain $\bar{\varepsilon}$, a quantity defined, in the simplest case, as a weighted spatial average of the equivalent strain $\widetilde\varepsilon$:
\begin{equation}
\bar\varepsilon(\bm x):=\displaystyle\frac{1}{|V|}\int_V g(\bm\xi)\widetilde\varepsilon(\bm x+\bm\xi) \,\mathrm{d}\bm\xi.
\label{EquivStr}
\end{equation}
In \autoref{EquivStr}, $V\subset\Omega$ is an averaging volume, $\bm\xi\in V$, and $g$ is a bell-shaped function, typically a Gaussian, $g(\bm\xi):=A\exp({-|\bm\xi|^2/(2l_\text{c}^2)})$, with $l_\text{c}$ related to the size (in this case, the radius) of $V$. Parameter $l_\text{c}$ may have a physical meaning: it can represent the characteristic length of the non-local continuum, and enable the description of strain localisation and size effects, see \cite{Peerl,Parisio2018} for further discussion. Finally, the normalization constant $A$ is chosen to provide $\frac{1}{|V|}\int_V g(\bm\xi)\,\mathrm{d}\bm\xi=1$. 

The described procedure is typically called regularization of the local CDM problem and results in what is termed a {\em non-local} formulation \cite{PijCabBaz1,Mazars2}. With $\bar\varepsilon$ at hand, such that now $D=D(\bar\varepsilon)$, the formulation reads:
\begin{equation}
\left\{
\begin{tabular}{c}
$\bm\sigma(\bm u,D):= \bigl(1-D(\bar{\varepsilon})\bigr)\,\mathbb{C}:\bm\varepsilon$, \\[0.2cm]
$\bar{\varepsilon}$ given by \autoref{EquivStr} or similar,
\end{tabular}
\right.
\label{NonLocalCMD}
\end{equation}
and can be shown to be free of the pathological mesh dependence of the original local one. 

One can notice that \autoref{NonLocalCMD} is an integro-differential equation system. To make its numerical treatment more straightforward, an idea -- a simple, but very elegant one -- of recasting the integral in \autoref{EquivStr} as a differential equation has been proposed. This eventually resulted in what is nowadays called a {\em gradient-enhanced} CDM formulation. For the sake of consistency, the technical derivation steps leading to the formulation are briefly outlined\footnote{That also gives a useful hint on how convolution integrals can, in general, be further on elaborated, if/when such a necessity occurs.}.

First, the integrand in the convolution integral \autoref{EquivStr} is expanded using a Taylor series (for the sake of simplicity, we consider and present the 2-dimensional case):
\begin{align}
    \widetilde\varepsilon(\bm x+\bm\xi)
    &=\widetilde\varepsilon(\bm x)
    +\frac{\partial\widetilde\varepsilon}{\partial x_1}(\bm x)\xi_1
    +\frac{\partial\widetilde\varepsilon}{\partial x_2}(\bm x)\xi_2 +\nonumber \\
    &+\frac{1}{2}\frac{\partial^2\widetilde\varepsilon}{\partial x_1^2}(\bm x)\xi_1^2
    +\frac{\partial^2\widetilde\varepsilon}{\partial x_1\partial x_2}(\bm x)\xi_1\xi_2
    +\frac{1}{2}\frac{\partial^2\widetilde\varepsilon}{\partial x_2^2}(\bm x)\xi_2^2+\mbox{h.o.t.}
\label{Taylor}. 
\end{align}
with h.o.t.\ denoting the higher order derivatives in $\bm x$, as well as the higher polynomial degrees of $\bm\xi$. Then, substituting \autoref{Taylor} in \autoref{EquivStr} and exployting some convenient properties the function $g$ possesses, namely,
\begin{equation}
\int_V g(\bm\xi)\xi_i \,\mathrm{d}\bm\xi=0, \; i=1,2, \;\;
\int_V g(\bm\xi)\xi_1\xi_2 \,\mathrm{d}\bm\xi=0,
\;\; \mbox{and} \;\;
\int_V g(\bm\xi)\xi_1^2 \; d\bm\xi=\int_V g(\bm\xi)\xi_2^2 \,\mathrm{d}\bm\xi,
\label{NicePrts}
\end{equation}
one obtains $\bar{\varepsilon}(\bm x) =\widetilde{\varepsilon}(\bm x) + c\Delta\widetilde{\varepsilon}(\bm x) + O(l_\text{c}^3)$ with $\Delta$ being the Laplace operator and $c\sim l_\text{c}^2$. The term $O(l_\text{c}^3)$ can be neglected, and we obtain the differential approximation of the integral relation \autoref{EquivStr}:
\begin{equation}
\bar{\varepsilon}(\bm x) =\widetilde{\varepsilon}(\bm x) + c\Delta\widetilde{\varepsilon}(\bm x).
\label{GradTerm}
\end{equation}
Applying the Laplace operator to \autoref{GradTerm}, multiplying it by $c$ and subtracting the result from \autoref{GradTerm}, one arrives at relation $-c\Delta\bar{\varepsilon}+\bar{\varepsilon}=\widetilde{\varepsilon}+\mathrm{h.o.t}$. Finally, neglecting in the above the terms of order four and higher, the desired gradient-enhanced formulation of \autoref{NonLocalCMD} by Peerlings et al. \cite{Peerl} is found:
\begin{equation}
\left\{
\begin{tabular}{l}
$\bm\sigma(\bm u,D):= \bigl(1-D(\bar{\varepsilon})\bigr)\,\mathbb{C}:\bm\varepsilon$, \\[0.2cm]
$-c\Delta\bar{\varepsilon}+\bar{\varepsilon}=\widetilde{\varepsilon}(\bm\varepsilon)$.
\end{tabular}
\right.
\label{GradEnh}
\end{equation}

The notion of ``gradient'' is suggested due to $\Delta=\nabla\cdot\nabla$. Note that in \autoref{GradEnh}, where $\bar{\varepsilon}$ is now an independent (extra) variable to be solved for and the right-hand side may be viewed as the driving force of its evolution. The numerical (finite element) treatment of \autoref{GradEnh} is much more straightforward than handling the original system \autoref{NonLocalCMD}. It can be also noticed that, by no surprise, in the limiting case $c\rightarrow0$ in \autoref{GradEnh}.b, the non-local equivalent strain turns into the local one, and the original local CDM formulation is recovered. This shows that the proposed regularization procedure is at least formally correct, since it preserves the so-called homotopic (that is, continuous) parametric link between the two formulations.

\subsection{Phase-field formulation of brittle fracture}

Another interesting non-standard instance of convolution can be found in the {\em phase-field modeling of fracture} (PMF). From the standpoint of material degradation modeling, this modern computational framework is somewhat similar to CDM, although the fundamentals, origins, capabilities as well as the related finite element treatment are different, see the review paper \cite{Ambati2015review}. In what follows, we show that in the PMF case, the concept of convolution finds its place as well, yet---remarkably---it is not introduced for problem regularisation purposes. Instead, it may help justify (or, one can say, support) a mathematical model of a process which has been obtained in a non-variational way\footnote{The advantage of deriving a boundary value problem (BVP) for a PDE which describes a certain (mechanical) process using a {\em variational approach}, or {\em principle of stationary or minimum energy} is that one has at the disposal a powerful tool of functional analysis enabling to prove existence and uniqueness of solution of a problem. Such a finding is naturally very helpful when it comes to finite element implementations, since variational consistency of the underlying problem allows one to guarantee that the computed approximation is physically meaningful. If a BVP problem is phenomenological, such as, e.g., the previously presented CDM formulation, and hence typically lacks variational consistency, it may be difficult, if not impossible, to establish the solution existence and uniqueness result.}.

\begin{figure}[!ht]
\begin{center}
\includegraphics[width=1.0\textwidth]{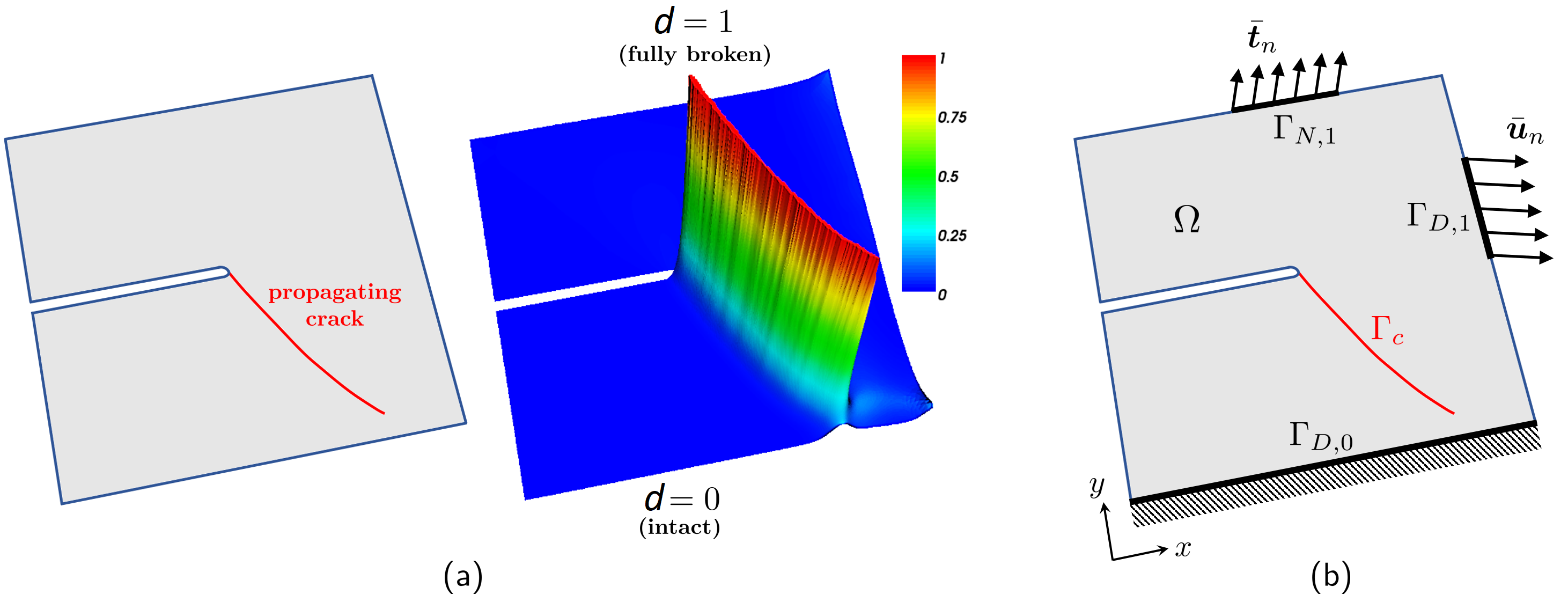}
\end{center}
\caption{(a) Phase-field description of fracture (sketch) with $d\in C(\Omega,[0,1])$ as the crack phase-field; (b) possible  mechanical system setup.}
\label{fig:PF}
\end{figure}

Thus, let us consider a process of deformation and fracture of a (brittle) elastic medium under quasi-static or dynamic external loading. Geometrically, by fracture we understand a crack (or multiple cracks), which in turn is a connected set of points where the material properties of the medium are discontinuous (the material is fully broken). In a phase-field formulation of fracture, a discrete crack with a property jump is represented in a smeared way: one introduces a variable $d$ which distinguishes between fully broken and intact material phases by taking the value 1 at a crack set, the value 0 where the material is intact, while it varies smoothly between both limiting values in a very ``thin'' transition zone, see \autoref{fig:PF}(a) for illustration. The spatial evolution of $d$ as a result of loading mimics the initiation and growth of a fracture.

In contrast to phenomenologically derived CDM models, the PMF formulation stems from a constrained minimization problem for a certain fracture energy functional. In a quasi-static setting, the generic formulation is given by the system of equations:
\begin{equation}
\left\{
\begin{tabular}{l}
$\bm\sigma(\bm u,d)=(1-d)^2\displaystyle\frac{\partial\Psi(\bm\varepsilon)}{\partial\bm\varepsilon}$, \\[0.2cm]
$-\ell^2\Delta d + d \geq \displaystyle\frac{2\ell}{G_\text{c}}(1-d)\Psi(\bm\varepsilon)$, \\[0.2cm]
$\dot{d}\geq0$, \\[0.2cm]
$\left(-\ell^2\Delta d + d - \displaystyle\frac{2\ell}{G_\text{c}}(1-d)\Psi(\bm\varepsilon)\right)\dot{d}=0$,
\end{tabular}
\right.
\label{Isotr}
\end{equation}
where $\Psi$ is the elastic free Helmholtz energy density function, $G_\text{c}$ is the material fracture toughness and $\ell$ is a (small) parameter which implicitly defines a thickness of a transtion zone of $d$. In the context of linear elasticity, $\Psi(\bm\varepsilon):=\frac{1}{2}\bm\varepsilon:\mathbb{C}:\bm\varepsilon$ thus yielding $\frac{\partial\Psi(\bm\varepsilon)}{\partial\bm\varepsilon}=\mathbb{C}:\bm\varepsilon$ and hence \autoref{Isotr}.a describes the material stifness degradation in the spirit of the CDM setting. The last three equations in \autoref{Isotr} describe the evolution of $d$: we have here differential inequality \autoref{Isotr}.b incorporating the crack phase-field driving force, a crack phase-field irreversibility constraint \autoref{Isotr}.c which is imposed to prevent crack healing and, finally, a complementary condition \autoref{Isotr}.d which relates the previous two. Clearly, these are more complicated than the Helmholz-type equation \autoref{GradEnh}.b for the non-local equivalent strain $\bar{\varepsilon}$ that drives the evolution of damage variable $D$.

For the numerical implementation, a simplified formulation stemming from \autoref{Isotr} has been proposed in \cite{Miehe1,Miehe2}:  
\begin{equation}
\left\{
\begin{tabular}{l}
$\bm\sigma(\bm u,d)=(1-d)^2\displaystyle\frac{\partial\Psi(\bm\varepsilon)}{\partial\bm\varepsilon}$, \\[0.2cm]
$-\ell^2\Delta d + d = \displaystyle\frac{2\ell}{G_\text{c}}(1-d)
\max_{\tau\in[0,t]}\Psi(\bm\varepsilon)$,
\end{tabular}
\right.
\label{IsotrHist}
\end{equation}
with $t$ standing for  pseudo-time. The second equation in \autoref{IsotrHist} has been formed using a reasonable phenomenological assumption about the role of the strain energy density $\Psi$ in \autoref{Isotr}.b as a driving force of the crack phase-field evolution. It substitutes relations \autoref{Isotr}.b--\autoref{Isotr}.d in an attempt to preserve and reproduce crack irreversibility. Unfortunately, \autoref{IsotrHist} is no longer of a variational nature and should be ``validated'', ideally not in a meerly computational way, but more rigorously. As it turns out, the concept of convolution provides such an option.

Thus, suppose $d:\Omega\rightarrow\mathbb{R}$ is a solution of \autoref{IsotrHist}.b. We want to show that there exist (scalar-valued) functions $F(s)$ and $\Phi(\ell;\bm x)$ such that assuming the representation
\begin{equation}
d(\bm x):=F\left( \frac{1}{|V|}\int_V g(\bm\xi)\Phi(\ell;\bm x+\bm\xi)\,\mathrm{d}\bm\xi \right),
\label{OurD}
\end{equation}
where $V\subset\Omega$ has the size of $\ell$ and $g(\bm\xi):=A\exp({-|\bm\xi|^2/(2\ell^2)})$, one recovers \autoref{IsotrHist}.b).

\begin{figure}[!ht]
\begin{center}
\includegraphics[width=1.0\textwidth]{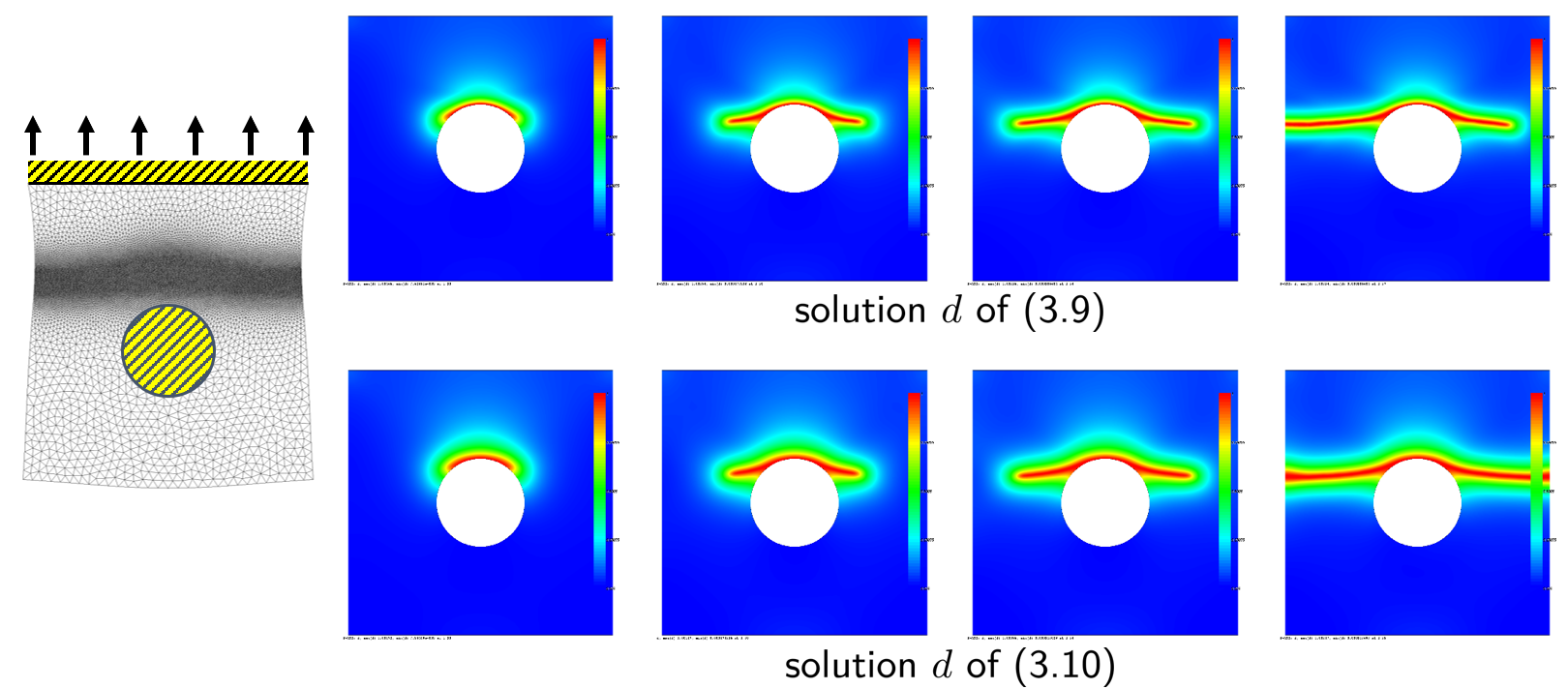}
\end{center}
\caption{Traction experiment on a fiber-reinforced matrix: crack phase-field evolution simulations using formulations \autoref{Isotr} and \autoref{IsotrHist}; the original figure is taken from \cite{LDLandTG2020}.}
\label{fig:PF_sim}
\end{figure}

The derivation of $F$ and $\Phi$ for \autoref{OurD} is rather straightforward. We start by employing the Taylor series expansion of $\Phi$ along with integration  while again making use of the properties of $g$ given in \autoref{NicePrts}. We thus obtain
\begin{equation*}
d(\bm x)=F\left( \Phi(\ell;\bm x) + \ell^2\Delta\Phi(\ell;\bm x)  + O(\ell^3) \right),
\end{equation*}
where $\ell^2:=\frac{1}{2|V|}\int_V g(\bm\xi)\xi_1^2\,\mathrm{d}\bm\xi$. Taylor series expansion of $F$ yields
\begin{equation*}
d(\bm x)=F\left(\Phi(\ell;\bm x)\right) + \ell^2 F^{\prime}\left(\Phi(\ell;\bm x)\right)\Delta\Phi(\ell;\bm x)  + O(\ell^3),
\end{equation*}
which, inserted into both sides of \autoref{IsotrHist}.b), results in
\begin{equation*}
- \ell^2\Delta d +d := F(\Phi) -\ell^2 F^{\prime\prime}(\Phi)|\nabla\Phi|^2 + O(\ell^3),
\end{equation*}
and
\begin{equation*}
\frac{2\ell}{G_\text{c}}(1-d)\max_{\tau\in[0,t]}\Psi := \frac{2\ell}{G_\text{c}}(1-F(\Phi))\max_{\tau\in[0,t]}\Psi + O(\ell^3),
\end{equation*}
respectively. The equation that links $F$ and $\Phi$ then reads
\begin{equation}
F(\Phi) -\ell^2 F^{\prime\prime}(\Phi)|\nabla\Phi|^2 + O(\ell^3)
 = \frac{2\ell}{G_\text{c}}(1-F(\Phi))\max_{\tau\in[0,t]}\Psi + O(\ell^3).
\label{Step4link}
\end{equation}
Imposing $F^{\prime\prime}(s)\equiv0$ on $F$ (this is done in order to get rid off the $|\nabla\Phi|^2$ term), we obtain that $F^{\prime}(s)=c\in\mathbb{R}$ and $F(s)=cs$. Then, neglecting $O(\ell^3)$ and plugging the result for $F$ in \autoref{Step4link}, we arrive at
\begin{equation*}
\Phi = c^{-1}\frac{ \frac{2\ell}{G_\text{c}}\max_{\tau\in[0,t]}\Psi }
{ 1+\frac{2\ell}{G_\text{c}}\max_{\tau\in[0,t]}\Psi }.
\end{equation*}
The desired representation of $d$ satisfying \autoref{IsotrHist}.b is obtained: 
\begin{equation}
d(\bm x)=\frac{1}{|V|} \int_V g(\bm\xi) 
\frac{ \frac{2\ell}{G_\text{c}}\max_{\tau\in[0,t]}\Psi(\bm x+\bm\xi) }{ 1+\frac{2\ell}{G_\text{c}}\max_{\tau\in[0,t]}\Psi(\bm x+\bm\xi) }     
\,\mathrm{d}\bm\xi.
\label{OurDfinal}
\end{equation}
Note that convolution \autoref{OurDfinal} is physically meaningful, as it provides the generic properties of the crack phase-field variable $d$ at any $\bm x\in\Omega$: $d\rightarrow 1$ (totally broken phase) when $\Psi\rightarrow\infty$, and $d\rightarrow 0$ (undamaged phase) when $\Psi\rightarrow 0$. The former uses $\frac{1}{|V|}\int_V g(\bm\xi)\,\mathrm{d}\bm\xi=1$. Furthermore, it can be concluded that irreversibility of $d$ is fulfilled too, which was a major motivation in passing from system \autoref{Isotr}.b-\autoref{Isotr}.d to \autoref{IsotrHist}.b. As a result, the convolution seemed to help justify formulation \autoref{IsotrHist} which ceased to be variationally consistent. \autoref{fig:PF_sim} depicts simulated fracture evolution in a fiber-reinforced matrix subject to traction, obtained by solving the original PMF formulation \autoref{Isotr} and the related phenomenological modification \autoref{IsotrHist}. Note that the complete identity of the results cannot be expected given that the  formulations are different, as well as accounting for possible solution non-uniqueness phenomena associated to PMF energy functional non-convexity\footnote{In fact, a break of solution symmetry already observed in the upper right plot in the corresponding case is the manifisation of the mentioned phenomenon of non-unique solutions. We refer the interested reader to publication \cite{TG1,LDLandTG2020} where the topic is elaborated in more detail.}. The comparison, however, additionally to the presented above convolution-based procedure, provides the numerical evidence that such a modification is meaningful even in the absence of variational consistency.

\section{Conclusions}
\label{sec:conclusions}
The notions of non-locality and convolution can be encountered at all levels of engineering mechanics: from simple examples tought in undergraduate courses to advanced theories at the forefront of engineering research. The simple examples remain useful, however, as they are easily accessible, help form an intuitive picture and thus build trust in concepts which might otherwise remain obscure. We also saw how, once understood, different concepts can help elucidate physical interpretations of models and reveal new theoretical insights into existing theories. Finally, abstraction as an important scientific tool serves to find common ground for frameworks that may have their roots in seemingly unrelated concepts. Aside from classical verification and validation procedures, analyses of this type can be a great asset in building confidence in advanced engineering theories by lending additional theoretical support. We hope that the presentation was instructive and has sparked your interest in this broad topic. We once more wish to encourage you to dig deeper in the supplementary material, where you find more examples from other branches of engineering.


\section*{Acknowledgments} 
This work was supported by the German Research Foundation (DFG). The first author would also like to thank Holger Steeb and Francesco Parisio for interesting discussions on non-local phenomena on several occasions.


\appendix 
\clearpage

\section{Application-oriented introduction to convolution} %
\label{sec_convolution}
The convolution integral emerges naturally in linear differential equations, since it embodies the superposition principle.
Although it may appear difficult at first glance, it is probably the easiest way to capture non-local effects.
By the way, integral transforms, such as the Laplace-Transform (exponential functions as kernel), are convolutions as well.
We will give a brief definition of the convolution sum (discrete setting) and convolution integral (continuous setting) next, before we take a tour through different domains and emphasize the specific applications of convolution.

\subsection{Definition and properties} \label{sec_defprop}
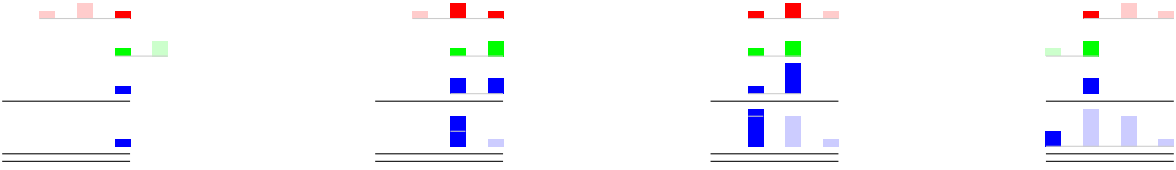
\begin{figure}
    \centering
\begin{tikzpicture}
\fill[red!20!white] (0, 0) +(-0.1, 0) rectangle +(0.1, 0.1);
\fill[red!20!white] (0.5, 0) +(-0.1, 0) rectangle +(0.1, 0.2);
\fill[red] (1.0, 0) +(-0.1, 0) rectangle +(0.1, 0.1);
\draw[thin, black!20!white] (-0.1, 0) -- (1.1, 0);
\fill[green] (1.0, -0.5) +(-0.1, 0) rectangle +(0.1, 0.1);
\fill[green!20!white] (1.5, -0.5) +(-0.1, 0) rectangle +(0.1, 0.2);
\draw[thin, black!20!white] (0.9, -0.5) -- (1.6, -0.5);
\fill[blue] (1, -1.0) +(-0.1, 0) rectangle +(0.1, 0.1);
\draw (-0.6, -1.1) -- (1.1, -1.1);
\fill[blue] (1, -1.7) +(-0.1, 0) rectangle +(0.1, 0.1);
\draw (-0.6, -1.8) -- (1.1, -1.8);
\draw (-0.6, -1.9) -- (1.1, -1.9);
\end{tikzpicture}
\hfill   
\begin{tikzpicture}
\fill[red!20!white] (0, 0) +(-0.1, 0) rectangle +(0.1, 0.1);
\fill[red] (0.5, 0) +(-0.1, 0) rectangle +(0.1, 0.2);
\fill[red] (1.0, 0) +(-0.1, 0) rectangle +(0.1, 0.1);
\draw[thin, black!20!white] (-0.1, 0) -- (1.1, 0);
\fill[green] (0.5, -0.5) +(-0.1, 0) rectangle +(0.1, 0.1);
\fill[green] (1.0, -0.5) +(-0.1, 0) rectangle +(0.1, 0.2);
\draw[thin, black!20!white] (0.4, -0.5) -- (1.1, -0.5);
\fill[blue] (0.5, -1.0) +(-0.1, 0) rectangle +(0.1, 0.2);
\fill[blue] (1, -1.0) +(-0.1, 0) rectangle +(0.1, 0.2);
\draw[thin, black!20!white] (0.4, -1.0) -- (1.1, -1.0);
\draw (-0.6, -1.1) -- (1.1, -1.1);
\fill[blue!20!white] (1, -1.7) +(-0.1, 0) rectangle +(0.1, 0.1);
\fill[blue] (0.5, -1.7) +(-0.1, 0) rectangle +(0.1, 0.4);
\draw[thin, black!20!white] (0.4, -1.5) -- (0.6, -1.5);
\draw (-0.6, -1.8) -- (1.1, -1.8);
\draw (-0.6, -1.9) -- (1.1, -1.9);
\end{tikzpicture}
\hfill   
\begin{tikzpicture}
\fill[red] (0, 0) +(-0.1, 0) rectangle +(0.1, 0.1);
\fill[red] (0.5, 0) +(-0.1, 0) rectangle +(0.1, 0.2);
\fill[red!20!white] (1.0, 0) +(-0.1, 0) rectangle +(0.1, 0.1);
\draw[thin, black!20!white] (-0.1, 0) -- (1.1, 0);
\fill[green] (0.0, -0.5) +(-0.1, 0) rectangle +(0.1, 0.1);
\fill[green] (0.5, -0.5) +(-0.1, 0) rectangle +(0.1, 0.2);
\draw[thin, black!20!white] (-0.1, -0.5) -- (0.6, -0.5);
\fill[blue] (0.0, -1.0) +(-0.1, 0) rectangle +(0.1, 0.1);
\fill[blue] (0.5, -1.0) +(-0.1, 0) rectangle +(0.1, 0.4);
\draw[thin, black!20!white] (-0.1, -1.0) -- (0.6, -1.0);
\draw (-0.6, -1.1) -- (1.1, -1.1);
\fill[blue!20!white] (1, -1.7) +(-0.1, 0) rectangle +(0.1, 0.1);
\fill[blue!20!white] (0.5, -1.7) +(-0.1, 0) rectangle +(0.1, 0.4);
\fill[blue] (0.0, -1.7) +(-0.1, 0) rectangle +(0.1, 0.5);
\draw[thin, black!20!white] (-0.1, -1.3) -- (0.1, -1.3);
\draw (-0.6, -1.8) -- (1.1, -1.8);
\draw (-0.6, -1.9) -- (1.1, -1.9);
\end{tikzpicture}
\hfill   
\begin{tikzpicture}
\fill[red] (0, 0) +(-0.1, 0) rectangle +(0.1, 0.1);
\fill[red!20!white] (0.5, 0) +(-0.1, 0) rectangle +(0.1, 0.2);
\fill[red!20!white] (1.0, 0) +(-0.1, 0) rectangle +(0.1, 0.1);
\draw[thin, black!20!white] (-0.1, 0) -- (1.1, 0);
\fill[green!20!white] (-0.5, -0.5) +(-0.1, 0) rectangle +(0.1, 0.1);
\fill[green] (0.0, -0.5) +(-0.1, 0) rectangle +(0.1, 0.2);
\draw[thin, black!20!white] (-0.6, -0.5) -- (0.1, -0.5);
\fill[blue] (0.0, -1.0) +(-0.1, 0) rectangle +(0.1, 0.2);
\draw (-0.6, -1.1) -- (1.1, -1.1);
\fill[blue!20!white] (1, -1.7) +(-0.1, 0) rectangle +(0.1, 0.1);
\fill[blue!20!white] (0.5, -1.7) +(-0.1, 0) rectangle +(0.1, 0.4);
\fill[blue!20!white] (0.0, -1.7) +(-0.1, 0) rectangle +(0.1, 0.5);
\fill[blue] (-0.5, -1.7) +(-0.1, 0) rectangle +(0.1, 0.2);
\draw[thin, black!20!white] (-0.6, -1.7) -- (1.1, -1.7);
\draw (-0.6, -1.8) -- (1.1, -1.8);
\draw (-0.6, -1.9) -- (1.1, -1.9);
\end{tikzpicture}
    \caption{Visualization of the polynomial multiplication $(x^2+2x+1)$\-$(2x+1)$\-$=2x^3+5x^2+4x+1$, the first coefficient vector (red) is fixed while the second (green) is flipped and shifted, for each shift are the element-wise products (blue, single underlined) summed and give a coefficient of the product (blue, double underlined).}
    \label{fig_dconvolution}
\end{figure}
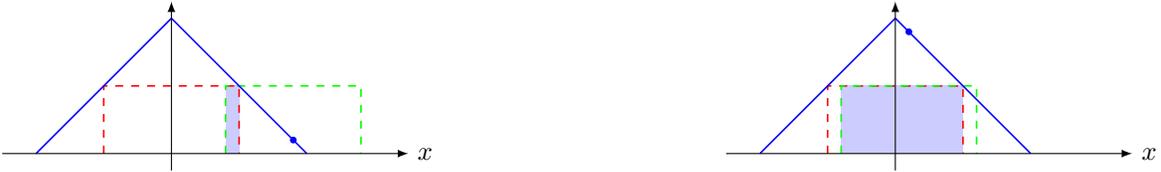
\begin{figure}
    \centering
\begin{tikzpicture}[scale=0.9]
\coordinate (A) at (1.8,0);   
\coordinate (B) at (1.8,0.2);   
\fill[blue!20!white] (A) +(-1, 0) rectangle (1, 1);
\draw[semithick,dashed,red] (-1,0) -- (-1,1) -- (1,1) -- (1,0);
\draw[semithick,dashed,green] (A) +(-1,0) -- +(-1,1) -- +(1,1) -- +(1,0);
\draw[semithick,blue] (-2, 0) -- (0, 2) -- (2,0);
\fill[blue] (B) circle[radius=0.05];
\draw[-latex] (-2.5, 0) -- (3.5, 0) node[right]{$x$};
\draw[-latex] ( 0,-.25) -- ( 0, 2.25);
\end{tikzpicture}
\hfill
\begin{tikzpicture}[scale=0.9]
\coordinate (A) at (0.2,0);   
\coordinate (B) at (0.2,1.8);   
\fill[blue!20!white] (A) +(-1, 0) rectangle (1, 1);
\draw[semithick,dashed,red] (-1,0) -- (-1,1) -- (1,1) -- (1,0);
\draw[semithick,dashed,green] (A) +(-1,0) -- +(-1,1) -- +(1,1) -- +(1,0);
\draw[semithick,blue] (-2, 0) -- (0, 2) -- (2,0);
\fill[blue] (B) circle[radius=0.05];
\draw[-latex] (-2.5, 0) -- (3.5, 0) node[right]{$x$};
\draw[-latex] ( 0,-.25) -- ( 0, 2.25);
\end{tikzpicture}
    \caption{Two sample points of the convolution (blue) of two rectangle functions (red, green), the blue dot corresponds to the shaded area.}
    \label{fig_convolution}
\end{figure}
A convolution is the superposition of the product of two functions after one of them is flipped and shifted. As such it can be considered as a particular kind of transform.
In the domain of (square-summable) discrete sequences it is defined by the sum
\begin{equation}
\label{eq_def_dc}
    F_3[n] = \sum_{k=-\infty}^{\infty} F_1[k] F_2[n-k]  \qquad \text{with} \quad n,k\in \mathbb{Z}.
\end{equation}
Alternatively, you may think of an index $l=n-k$ with the property $k+l=n$.

In the domain of (square-integrable) real-valued functions (straightforwardly generalizable to complex-valued) it is defined by the integral \cite{bronstein2012taschenbuch}
\begin{equation}
\label{eq_def_cc}
    f_3(x) = \int_{-\infty}^{\infty} f_1(\bar{x})f_2(x-\bar{x})\,\mathrm{d}\bar{x} \qquad \text{with} \quad x,\bar{x}\in \mathbb{R}. 
\end{equation}
Introducing the convolution operator (for discrete as continuous), we may write equations~\eqref{eq_def_dc} and \eqref{eq_def_cc}
\begin{align}
 F_3[n] &=  F_1[n] *  F_2[n], \\
 f_3(t) &=  f_1(t) *  f_2(t). 
\end{align}
Discrete convolutions you may have unconsciously met when multiplying polynomials. 
The coefficent vector of the product is the convolution of the coefficient vectors of the factors, for example
\begin{subequations}
\begin{align}
(x^2 + 2x + 1)(2x + 1)&= 2x^3 + 5x^2  + 4x + 1,\\
\bigl[\begin{array}{ccc} 1 &2& 1 \end{array}\bigr]
*\bigl[\begin{array}{cc} 2 & 1 \end{array}\bigr] 
&= \bigl[\begin{array}{cccc} 2 & 5 & 4 & 1 \end{array}\bigr],
\end{align}
\end{subequations}
which is illustrated in figure~\ref{fig_dconvolution}.
Convolutions of continuous functions is illustrated at the simple example of two rectangular functions, also known as boxcar, in figure~\ref{fig_convolution}.
In applications, one of the input functions characterizes the system, e.g. elastic half-space, and the other some excitation, e.g. the surface force distribution.
Then the output function corresponds to the system response, e.g. stress-field in the elastic-half space under given force surface force distribution (load).

We briefly summarize the important properties of commutativity, associativity and distributivity (over addition) \cite{bronstein2012taschenbuch} 
\begin{align}
 a*b &= b*a, \\
 a*(b*c) &= (a*b)*c, \\
 a*(b+c) &= a*b + a*c,
 \label{eq_distributivity}
\end{align}
where $a$, $b$, $c$ may represent either discrete sequences or continuous functions.
We also have associativity with scalar multiplication \cite{bronstein2012taschenbuch}, here with factor $\gamma$
\begin{equation}
 \gamma (a*b) = (\gamma a)*b = a*(\gamma b),
\end{equation}
which together with distributivity~\eqref{eq_distributivity} renders convolution a linear operation.

To complement our explanations, we refer to some very instructive visualizations on YouTube\footnote{\href{https://www.youtube.com/watch?v=KuXjwB4LzSA}{3Blue1Brown: But what is a convolution?}}.

\subsection{Statistical correlations}
Convolutions appear almost verbatim, except the sign of the shift, in statistics as auto- and cross-correlation
which measure the similarity of a signal with its time-shifted version or with another signal, respectively \cite{stengel2012optimal}.
In the domain of real numbers, these correlations read for discrete-time series $X[n]$ and $Y[n]$
\begin{subequations}
 \begin{align}
  R_{xx}[n] &= \sum_{m=-\infty}^{\infty} X[m]X[m+n] ,\\
  R_{xy}[n] &= \sum_{m=-\infty}^{\infty} X[m]Y[m+n] ,
 \end{align}
\end{subequations}
and for continuous-time functions $x(t)$ and $y(t)$
\begin{subequations}
 \begin{align}
R_{xx}(\tau) &= \int_{-\infty}^{\infty} x(t)x(t+\tau)\,\text{d}t ,\\
R_{xy}(\tau) &= \int_{-\infty}^{\infty} x(t)y(t+\tau)\,\text{d}t .
 \end{align}
\end{subequations}

\subsection{Sums of random variables}
We start with discrete distributions in which one has a finite set of possible results and then generalize to continuous distributions.
Presuming probabilities of supply and demand, we are interested in the probabilities of the surplus, defined as the difference between supply and demand.
As example for both kinds of distributions, we imagine an oral exam.
\textsl{Supply}, in this context, means the knowledge level of the student as a percentage of the curriculum, whereas \textsl{demand} signifies the expectation of the examiner.
We just refer to the colloquial meaning of \textsl{expectation}, not to be mistaken with the mathematically defined expected value.
The \textsl{surplus} measures the outcome of the exam, positive means passed: the higher the value of the surplus, the more happiness on both sides. 
Negative values mean the student failed the exam and the value measures frustration. 

\subsubsection{Discrete distributions}
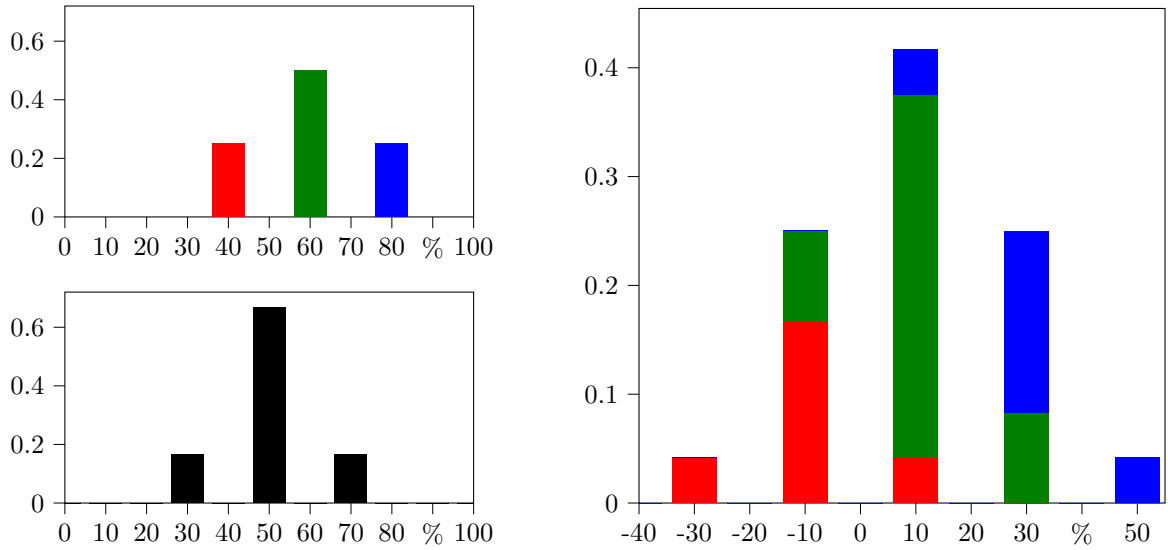
\begin{figure}
\setlength{\figW}{0.45\textwidth}
\setlength{\figH}{0.2\textheight}
\begin{tikzpicture}

\begin{groupplot}[group style={group size=1 by 2}]
\nextgroupplot[
height=\figH,
tick align=outside,
tick pos=left,
width=\figW,
x grid style={white!69.0196078431373!black},
xmin=0, xmax=10,
xtick style={color=black},
xtick={0,1,2,3,4,5,6,7,8,9,10},
xticklabels={0,10,20,30,40,50,60,70,80,\%,100},
y grid style={white!69.0196078431373!black},
ymin=0, ymax=0.72,
ytick style={color=black}
]
\draw[draw=none,fill=red] (axis cs:3.6,0) rectangle (axis cs:4.4,0.25);
\draw[draw=none,fill=green!50.1960784313725!black] (axis cs:5.6,0) rectangle (axis cs:6.4,0.5);
\draw[draw=none,fill=blue] (axis cs:7.6,0) rectangle (axis cs:8.4,0.25);

\nextgroupplot[
height=\figH,
tick align=outside,
tick pos=left,
width=\figW,
x grid style={white!69.0196078431373!black},
xmin=0, xmax=10,
xtick style={color=black},
xtick={0,1,2,3,4,5,6,7,8,9,10},
xticklabels={0,10,20,30,40,50,60,70,80,\%,100},
y grid style={white!69.0196078431373!black},
ymin=0, ymax=0.72,
ytick style={color=black}
]
\draw[draw=none,fill=black] (axis cs:-0.4,0) rectangle (axis cs:0.4,0);
\draw[draw=none,fill=black] (axis cs:0.6,0) rectangle (axis cs:1.4,0);
\draw[draw=none,fill=black] (axis cs:1.6,0) rectangle (axis cs:2.4,0);
\draw[draw=none,fill=black] (axis cs:2.6,0) rectangle (axis cs:3.4,0.166666666666667);
\draw[draw=none,fill=black] (axis cs:3.6,0) rectangle (axis cs:4.4,0);
\draw[draw=none,fill=black] (axis cs:4.6,0) rectangle (axis cs:5.4,0.666666666666667);
\draw[draw=none,fill=black] (axis cs:5.6,0) rectangle (axis cs:6.4,0);
\draw[draw=none,fill=black] (axis cs:6.6,0) rectangle (axis cs:7.4,0.166666666666667);
\draw[draw=none,fill=black] (axis cs:7.6,0) rectangle (axis cs:8.4,0);
\draw[draw=none,fill=black] (axis cs:8.6,0) rectangle (axis cs:9.4,0);
\draw[draw=none,fill=black] (axis cs:9.6,0) rectangle (axis cs:10.4,0);
\end{groupplot}

\end{tikzpicture}
\hfill
\setlength{\figW}{0.55\textwidth}
\setlength{\figH}{0.372\textheight}
\begin{tikzpicture}

\begin{axis}[
height=\figH,
tick align=outside,
tick pos=left,
width=\figW,
x grid style={white!69.0196078431373!black},
xmin=1, xmax=10.5,
xtick style={color=black},
xtick={1,2,3,4,5,6,7,8,9,10},
xticklabels={-40,-30,-20,-10,0,10,20,30,\%,50},
y grid style={white!69.0196078431373!black},
ymin=0, ymax=0.4545,
ytick style={color=black}
]
\draw[draw=none,fill=red] (axis cs:-0.4,0) rectangle (axis cs:0.4,0);
\draw[draw=none,fill=red] (axis cs:0.6,0) rectangle (axis cs:1.4,0);
\draw[draw=none,fill=red] (axis cs:1.6,0) rectangle (axis cs:2.4,0.0416666666666667);
\draw[draw=none,fill=red] (axis cs:2.6,0) rectangle (axis cs:3.4,0);
\draw[draw=none,fill=red] (axis cs:3.6,0) rectangle (axis cs:4.4,0.166666666666667);
\draw[draw=none,fill=red] (axis cs:4.6,0) rectangle (axis cs:5.4,0);
\draw[draw=none,fill=red] (axis cs:5.6,0) rectangle (axis cs:6.4,0.0416666666666667);
\draw[draw=none,fill=red] (axis cs:6.6,0) rectangle (axis cs:7.4,0);
\draw[draw=none,fill=red] (axis cs:7.6,0) rectangle (axis cs:8.4,0);
\draw[draw=none,fill=red] (axis cs:8.6,0) rectangle (axis cs:9.4,0);
\draw[draw=none,fill=red] (axis cs:9.6,0) rectangle (axis cs:10.4,0);
\draw[draw=none,fill=green!50.1960784313725!black] (axis cs:-0.4,0) rectangle (axis cs:0.4,0);
\draw[draw=none,fill=green!50.1960784313725!black] (axis cs:0.6,0) rectangle (axis cs:1.4,0);
\draw[draw=none,fill=green!50.1960784313725!black] (axis cs:1.6,0.0416666666666667) rectangle (axis cs:2.4,0.0416666666666667);
\draw[draw=none,fill=green!50.1960784313725!black] (axis cs:2.6,0) rectangle (axis cs:3.4,0);
\draw[draw=none,fill=green!50.1960784313725!black] (axis cs:3.6,0.166666666666667) rectangle (axis cs:4.4,0.25);
\draw[draw=none,fill=green!50.1960784313725!black] (axis cs:4.6,0) rectangle (axis cs:5.4,0);
\draw[draw=none,fill=green!50.1960784313725!black] (axis cs:5.6,0.0416666666666667) rectangle (axis cs:6.4,0.375);
\draw[draw=none,fill=green!50.1960784313725!black] (axis cs:6.6,0) rectangle (axis cs:7.4,0);
\draw[draw=none,fill=green!50.1960784313725!black] (axis cs:7.6,0) rectangle (axis cs:8.4,0.0833333333333333);
\draw[draw=none,fill=green!50.1960784313725!black] (axis cs:8.6,0) rectangle (axis cs:9.4,0);
\draw[draw=none,fill=green!50.1960784313725!black] (axis cs:9.6,0) rectangle (axis cs:10.4,0);
\draw[draw=none,fill=blue] (axis cs:-0.4,0) rectangle (axis cs:0.4,0);
\draw[draw=none,fill=blue] (axis cs:0.6,0) rectangle (axis cs:1.4,0);
\draw[draw=none,fill=blue] (axis cs:1.6,0.0416666666666667) rectangle (axis cs:2.4,0.0416666666666667);
\draw[draw=none,fill=blue] (axis cs:2.6,0) rectangle (axis cs:3.4,0);
\draw[draw=none,fill=blue] (axis cs:3.6,0.25) rectangle (axis cs:4.4,0.25);
\draw[draw=none,fill=blue] (axis cs:4.6,0) rectangle (axis cs:5.4,0);
\draw[draw=none,fill=blue] (axis cs:5.6,0.375) rectangle (axis cs:6.4,0.416666666666667);
\draw[draw=none,fill=blue] (axis cs:6.6,0) rectangle (axis cs:7.4,0);
\draw[draw=none,fill=blue] (axis cs:7.6,0.0833333333333333) rectangle (axis cs:8.4,0.25);
\draw[draw=none,fill=blue] (axis cs:8.6,0) rectangle (axis cs:9.4,0);
\draw[draw=none,fill=blue] (axis cs:9.6,0) rectangle (axis cs:10.4,0.0416666666666667);
\end{axis}

\end{tikzpicture}
\caption{Discrete probabilities for student's knowledge level (left top), examiners expectation (left bottom) and exam outcome (right).}
\label{fig_discrete_random}
\end{figure}
\begin{table}[htbp]
\centering
\footnotesize
\caption{Definition and probabilities of exemplary exam (discrete distribution)}
\label{tab_dexam}  
  \begin{tabular}{|lcc|lcc|} \hline
  \bf student & \bf knowledge & \bf probability & 
  \bf examiner & \bf expectation & \bf probability \\ \hline
    good & $\SI{80}{\percent}$ & $1/4$ &
    strict & $\SI{70}{\percent}$ & $1/6$ \\
    average & $\SI{60}{\percent}$ & $1/2$ & 
    average & $\SI{50}{\percent}$ & $2/3$ \\
    bad & $\SI{40}{\percent}$ & $1/4$ &
    laid-back & $\SI{30}{\percent}$ & $1/6$ \\ \hline 
  \end{tabular}
\end{table}
In a discrete setting we assume three kinds of students, and three kinds of examiners  with the numbers shown in \autoref{tab_dexam}.
Let's start with a good student, depending on the examiner the result will be either $\SI{50}{\percent}$, $\SI{30}{\percent}$ or $\SI{10}{\percent}$ above the expectation. 
Similarly, for an average student the result will be either $\SI{30}{\percent}$, $\SI{10}{\percent}$ or $\SI{-10}{\percent}$. 
For a bad student, the result will be $\SI{10}{\percent}$, $\SI{-10}{\percent}$ or $\SI{-30}{\percent}$.
Note, that there are two possibilities for an outcome of $\SI{30}{\percent}$, a good student meets an average examiner or an average student meets a laid-back examiner. 
For an outcome of $\SI{10}{\percent}$ there will be three possibilities and for $\SI{-10}{\percent}$ two possibilities.
This is where the convolution comes in, for each student we obtain an outcome distribution which we have to add together with the other students
\begin{equation}
\label{eq_dd_sum}
P_\text{surplus}(\Delta) = \sum_{n=1}^{N_\text{supply}} P_\text{supply}(n) P_\text{demand}(n-\Delta).
\end{equation}
The resulting probabilities for exam outcome and its composition are illustrated in \autoref{fig_discrete_random}.
Convolution here means the summation over the probabilities of supply distributed over the demand.
In this example it blends students knownledge with examiners expectation.
As you may easily verify by a tree diagram, there will be the same outcome when you swap the order and firstly select the examiner and secondly the student, since the convolution is a commutative operation.

\subsubsection{Continuous distributions}
\begin{figure}
\centering
\setlength{\figW}{0.6\textwidth}
\setlength{\figH}{0.4\textwidth}
\input{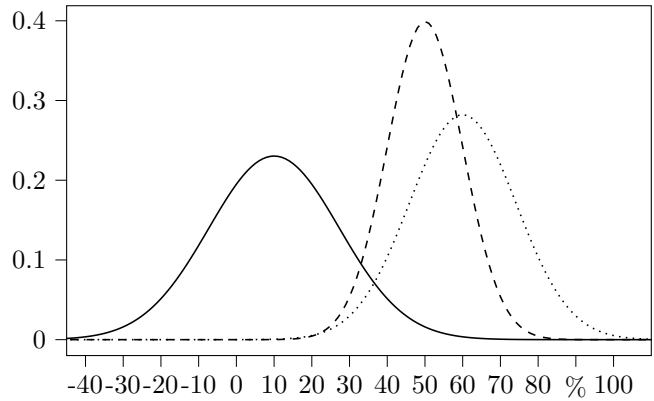}
\caption{Probability density functions of student's knowledge (dotted line), examiner's expectation (dashed line) and exam outcome (solid line).}
\label{fig_continuous_random}
\end{figure}
In the continuous setting there are no more finite numbers of students and examiners with a certain probability, but we transition from histogrammes to probability density functions. 
Assuming normal distributions, we may describe the student population's knowledge by a mean value $\mu_\text{S}=\SI{60}{\percent}$ and variance $\sigma_\text{S}^2=\SI{20}{\percent}$.
Similarly, we assume for the examiner population $\mu_\text{E}=\SI{50}{\percent}$ and $\sigma_\text{E}^2=\SI{10}{\percent}$.
Now instead of the sum \eqref{eq_dd_sum} we need to evaluate the integral
\begin{equation}
\label{eq_prob_conv}
 p_\text{surplus}(\Delta) = \int_{-\infty}^\infty p_\text{supply}(x) p_\text{demand}(x-\Delta)\,\text{d}x.
\end{equation}
Figure~\ref{fig_continuous_random} shows all three probability density functions.
Assuming independent normal distributions in equation~\eqref{eq_prob_conv} we obtain also a normal distribution as result with
\begin{align}
 \mu_\text{surplus} &= \mu_\text{supply} - \mu_\text{demand}, \\
 \sigma_\text{surplus}^2 &= \sigma_\text{supply}^2 + \sigma_\text{demand}^2.
\end{align}
For our values they are $\mu_\text{surplus}=\SI{10}{\percent}$ and $\sigma_\text{surplus}^2=\SI{30}{\percent}$.
For a sum instead of a difference you would have to add the mean values, whereas the resulting variance is always a summation.

\textsl{Local} would mean in this context that either all examiners have the same expectation (ideal case) or there are uniform students (strange case), similarly to the simplication \ref{eq:fl} in Boussinesq's problem.

Another scenario of this type you will find in probabilistic design.
There surplus means safety and is defined as difference between strength and load. Departing from this, reliability indices and failure properties can be calculated.

\subsection{Linear time-invariant systems}
\begin{figure}
\centering
\begin{tikzpicture}
\draw[thick, -latex] (-3.9, 0) -- node[above]{$u_1$} (-3, 0);
\draw[semithick] (-3,-1) rectangle (-1, 1);
\draw (-2, 0) node {$H_1$};
\draw[thick, -latex] (-1, 0) -- node[above]{$y_1$} (-0.1, 0);
\draw[thick, -latex] ( 0.1, 0) -- node[above]{$u_2$} ( 1, 0);
\draw[semithick] ( 1,-1) rectangle ( 3, 1);
\draw ( 2, 0) node {$H_2$};
\draw[thick, -latex] ( 3, 0) -- node[above]{$y_2$} ( 3.9, 0);
\draw[dashed, semithick] (-4,-1.25) rectangle (4, 1.25);
\draw[thick, -latex] (-5.0, 0) -- node[above] {$u_{12}$} (-4.1, 0);
\draw[thick, -latex] ( 4.1, 0) -- node[above] {$y_{12}$} ( 5.0, 0);
\end{tikzpicture} 
\caption{Series connection of two LTI-systems with transfer functions $H_1$ and $H_2$, respectively; $u_{12}$ is the input to the compound and $y_{12}$ the corresponding output signal.}
\label{fig_series_connection}
\end{figure}
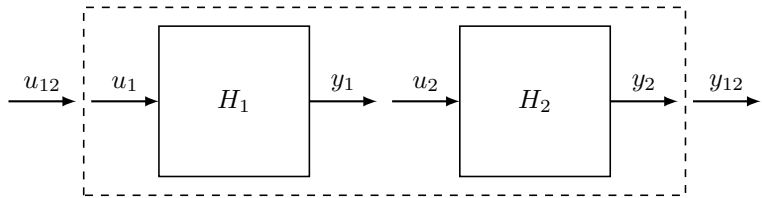
Convolution plays a fundamental role in signal processing and control theory for linear systems.
We demonstrate its use to compute the response of linear time-invariant systems.
Since in a series connection (figure~\ref{fig_series_connection}) the output of the first system is
the input to the second system, convolution helps us to find the transfer function of the compound.

As previously, 
we start with a discrete setting and then generalize to the continuous setting.

\subsubsection{Discrete-time LTI-systems}
\begin{figure}
\setlength{\figW}{0.40\textwidth}
\setlength{\figH}{0.20\textheight}
\begin{tabular}{cccc}
& averaging filter & backward difference & series connection \\
\begin{turn}{90} \hspace{0.05\textheight} $u[k]$ \end{turn} & 
\begin{tikzpicture}

\definecolor{darkgray176}{RGB}{176,176,176}

\begin{axis}[
height=\figH,
hide x axis,
hide y axis,
tick align=outside,
tick pos=left,
width=\figW,
x grid style={darkgray176},
xmin=-1.1, xmax=12.1,
xtick style={color=black},
y grid style={darkgray176},
ymin=-1.1, ymax=1.1,
ytick style={color=black}
]
\path [draw=white, semithick]
(axis cs:0,0)
--(axis cs:0,-1);

\path [draw=white, semithick]
(axis cs:1,0)
--(axis cs:1,1);

\addplot [semithick, white, mark=*, mark size=3, mark options={solid}, only marks]
table {%
0 -1
1 1
};
\addplot [semithick, white]
table {%
0 0
1 0
};
\path [draw=black, semithick]
(axis cs:0,0)
--(axis cs:0,0);

\path [draw=black, semithick]
(axis cs:1,0)
--(axis cs:1,0);

\path [draw=black, semithick]
(axis cs:2,0)
--(axis cs:2,0);

\path [draw=black, semithick]
(axis cs:3,0)
--(axis cs:3,0.333333333333333);

\path [draw=black, semithick]
(axis cs:4,0)
--(axis cs:4,0.666666666666667);

\path [draw=black, semithick]
(axis cs:5,0)
--(axis cs:5,0.666666666666667);

\path [draw=black, semithick]
(axis cs:6,0)
--(axis cs:6,0.666666666666667);

\path [draw=black, semithick]
(axis cs:7,0)
--(axis cs:7,0.333333333333333);

\path [draw=black, semithick]
(axis cs:8,0)
--(axis cs:8,0);

\path [draw=black, semithick]
(axis cs:9,0)
--(axis cs:9,0);

\path [draw=black, semithick]
(axis cs:10,0)
--(axis cs:10,0);

\path [draw=black, semithick]
(axis cs:11,0)
--(axis cs:11,0);

\addplot [semithick, black, mark=*, mark size=3, mark options={solid}, only marks]
table {%
0 0
1 0
2 0
3 0.333333333333333
4 0.666666666666667
5 0.666666666666667
6 0.666666666666667
7 0.333333333333333
8 0
9 0
10 0
11 0
};
\addplot [black]
table {%
0 0
11 0
};
\path [draw=black, semithick]
(axis cs:0,0)
--(axis cs:0,0);

\addplot [semithick, black, mark=*, mark size=3, mark options={solid,fill=white}, only marks]
table {%
0 0
};
\addplot [semithick, black]
table {%
0 0
0 0
};
\end{axis}

\end{tikzpicture} & 
\begin{tikzpicture}

\definecolor{darkgray176}{RGB}{176,176,176}

\begin{axis}[
height=\figH,
hide x axis,
hide y axis,
tick align=outside,
tick pos=left,
width=\figW,
x grid style={darkgray176},
xmin=-1.1, xmax=12.1,
xtick style={color=black},
y grid style={darkgray176},
ymin=-1.1, ymax=1.1,
ytick style={color=black}
]
\path [draw=white, semithick]
(axis cs:0,0)
--(axis cs:0,-1);

\path [draw=white, semithick]
(axis cs:1,0)
--(axis cs:1,1);

\addplot [semithick, white, mark=*, mark size=3, mark options={solid}, only marks]
table {%
0 -1
1 1
};
\addplot [semithick, white]
table {%
0 0
1 0
};
\path [draw=black, semithick]
(axis cs:0,0)
--(axis cs:0,0);

\path [draw=black, semithick]
(axis cs:1,0)
--(axis cs:1,0);

\path [draw=black, semithick]
(axis cs:2,0)
--(axis cs:2,0);

\path [draw=black, semithick]
(axis cs:3,0)
--(axis cs:3,0.333333333333333);

\path [draw=black, semithick]
(axis cs:4,0)
--(axis cs:4,0.666666666666667);

\path [draw=black, semithick]
(axis cs:5,0)
--(axis cs:5,0.666666666666667);

\path [draw=black, semithick]
(axis cs:6,0)
--(axis cs:6,0.666666666666667);

\path [draw=black, semithick]
(axis cs:7,0)
--(axis cs:7,0.333333333333333);

\path [draw=black, semithick]
(axis cs:8,0)
--(axis cs:8,0);

\path [draw=black, semithick]
(axis cs:9,0)
--(axis cs:9,0);

\path [draw=black, semithick]
(axis cs:10,0)
--(axis cs:10,0);

\path [draw=black, semithick]
(axis cs:11,0)
--(axis cs:11,0);

\addplot [semithick, black, mark=*, mark size=3, mark options={solid}, only marks]
table {%
0 0
1 0
2 0
3 0.333333333333333
4 0.666666666666667
5 0.666666666666667
6 0.666666666666667
7 0.333333333333333
8 0
9 0
10 0
11 0
};
\addplot [black]
table {%
0 0
11 0
};
\path [draw=black, semithick]
(axis cs:0,0)
--(axis cs:0,0);

\addplot [semithick, black, mark=*, mark size=3, mark options={solid,fill=white}, only marks]
table {%
0 0
};
\addplot [semithick, black]
table {%
0 0
0 0
};
\end{axis}

\end{tikzpicture} & 
\begin{tikzpicture}

\definecolor{darkgray176}{RGB}{176,176,176}

\begin{axis}[
height=\figH,
hide x axis,
hide y axis,
tick align=outside,
tick pos=left,
width=\figW,
x grid style={darkgray176},
xmin=-1.1, xmax=12.1,
xtick style={color=black},
y grid style={darkgray176},
ymin=-1.1, ymax=1.1,
ytick style={color=black}
]
\path [draw=white, semithick]
(axis cs:0,0)
--(axis cs:0,-1);

\path [draw=white, semithick]
(axis cs:1,0)
--(axis cs:1,1);

\addplot [semithick, white, mark=*, mark size=3, mark options={solid}, only marks]
table {%
0 -1
1 1
};
\addplot [semithick, white]
table {%
0 0
1 0
};
\path [draw=black, semithick]
(axis cs:0,0)
--(axis cs:0,0);

\path [draw=black, semithick]
(axis cs:1,0)
--(axis cs:1,0);

\path [draw=black, semithick]
(axis cs:2,0)
--(axis cs:2,0);

\path [draw=black, semithick]
(axis cs:3,0)
--(axis cs:3,0.333333333333333);

\path [draw=black, semithick]
(axis cs:4,0)
--(axis cs:4,0.666666666666667);

\path [draw=black, semithick]
(axis cs:5,0)
--(axis cs:5,0.666666666666667);

\path [draw=black, semithick]
(axis cs:6,0)
--(axis cs:6,0.666666666666667);

\path [draw=black, semithick]
(axis cs:7,0)
--(axis cs:7,0.333333333333333);

\path [draw=black, semithick]
(axis cs:8,0)
--(axis cs:8,0);

\path [draw=black, semithick]
(axis cs:9,0)
--(axis cs:9,0);

\path [draw=black, semithick]
(axis cs:10,0)
--(axis cs:10,0);

\path [draw=black, semithick]
(axis cs:11,0)
--(axis cs:11,0);

\addplot [semithick, black, mark=*, mark size=3, mark options={solid}, only marks]
table {%
0 0
1 0
2 0
3 0.333333333333333
4 0.666666666666667
5 0.666666666666667
6 0.666666666666667
7 0.333333333333333
8 0
9 0
10 0
11 0
};
\addplot [black]
table {%
0 0
11 0
};
\path [draw=black, semithick]
(axis cs:0,0)
--(axis cs:0,0);

\addplot [semithick, black, mark=*, mark size=3, mark options={solid,fill=white}, only marks]
table {%
0 0
};
\addplot [semithick, black]
table {%
0 0
0 0
};
\end{axis}

\end{tikzpicture} \\
\begin{turn}{90} \hspace{0.05\textheight} $h[k]$ \end{turn}  & 
\begin{tikzpicture}

\definecolor{darkgray176}{RGB}{176,176,176}

\begin{axis}[
height=\figH,
hide x axis,
hide y axis,
tick align=outside,
tick pos=left,
width=\figW,
x grid style={darkgray176},
xmin=-1.1, xmax=12.1,
xtick style={color=black},
y grid style={darkgray176},
ymin=-1.1, ymax=1.1,
ytick style={color=black}
]
\path [draw=white, semithick]
(axis cs:0,0)
--(axis cs:0,-1);

\path [draw=white, semithick]
(axis cs:1,0)
--(axis cs:1,1);

\addplot [semithick, white, mark=*, mark size=3, mark options={solid}, only marks]
table {%
0 -1
1 1
};
\addplot [semithick, white]
table {%
0 0
1 0
};
\path [draw=black, semithick]
(axis cs:0,0)
--(axis cs:0,0.333333333333333);

\path [draw=black, semithick]
(axis cs:1,0)
--(axis cs:1,0.333333333333333);

\path [draw=black, semithick]
(axis cs:2,0)
--(axis cs:2,0.333333333333333);

\path [draw=black, semithick]
(axis cs:3,0)
--(axis cs:3,0);

\path [draw=black, semithick]
(axis cs:4,0)
--(axis cs:4,0);

\path [draw=black, semithick]
(axis cs:5,0)
--(axis cs:5,0);

\path [draw=black, semithick]
(axis cs:6,0)
--(axis cs:6,0);

\path [draw=black, semithick]
(axis cs:7,0)
--(axis cs:7,0);

\path [draw=black, semithick]
(axis cs:8,0)
--(axis cs:8,0);

\path [draw=black, semithick]
(axis cs:9,0)
--(axis cs:9,0);

\path [draw=black, semithick]
(axis cs:10,0)
--(axis cs:10,0);

\path [draw=black, semithick]
(axis cs:11,0)
--(axis cs:11,0);

\addplot [semithick, black, mark=*, mark size=3, mark options={solid}, only marks]
table {%
0 0.333333333333333
1 0.333333333333333
2 0.333333333333333
3 0
4 0
5 0
6 0
7 0
8 0
9 0
10 0
11 0
};
\addplot [black]
table {%
0 0
11 0
};
\path [draw=black, semithick]
(axis cs:0,0)
--(axis cs:0,0.333333333333333);

\addplot [semithick, black, mark=*, mark size=3, mark options={solid,fill=white}, only marks]
table {%
0 0.333333333333333
};
\addplot [semithick, black]
table {%
0 0
0 0
};
\end{axis}

\end{tikzpicture} & 
\begin{tikzpicture}

\definecolor{darkgray176}{RGB}{176,176,176}

\begin{axis}[
height=\figH,
hide x axis,
hide y axis,
tick align=outside,
tick pos=left,
width=\figW,
x grid style={darkgray176},
xmin=-1.1, xmax=12.1,
xtick style={color=black},
y grid style={darkgray176},
ymin=-1.1, ymax=1.1,
ytick style={color=black}
]
\path [draw=white, semithick]
(axis cs:0,0)
--(axis cs:0,-1);

\path [draw=white, semithick]
(axis cs:1,0)
--(axis cs:1,1);

\addplot [semithick, white, mark=*, mark size=3, mark options={solid}, only marks]
table {%
0 -1
1 1
};
\addplot [semithick, white]
table {%
0 0
1 0
};
\path [draw=black, semithick]
(axis cs:0,0)
--(axis cs:0,1);

\path [draw=black, semithick]
(axis cs:1,0)
--(axis cs:1,-1);

\path [draw=black, semithick]
(axis cs:2,0)
--(axis cs:2,0);

\path [draw=black, semithick]
(axis cs:3,0)
--(axis cs:3,0);

\path [draw=black, semithick]
(axis cs:4,0)
--(axis cs:4,0);

\path [draw=black, semithick]
(axis cs:5,0)
--(axis cs:5,0);

\path [draw=black, semithick]
(axis cs:6,0)
--(axis cs:6,0);

\path [draw=black, semithick]
(axis cs:7,0)
--(axis cs:7,0);

\path [draw=black, semithick]
(axis cs:8,0)
--(axis cs:8,0);

\path [draw=black, semithick]
(axis cs:9,0)
--(axis cs:9,0);

\path [draw=black, semithick]
(axis cs:10,0)
--(axis cs:10,0);

\path [draw=black, semithick]
(axis cs:11,0)
--(axis cs:11,0);

\addplot [semithick, black, mark=*, mark size=3, mark options={solid}, only marks]
table {%
0 1
1 -1
2 0
3 0
4 0
5 0
6 0
7 0
8 0
9 0
10 0
11 0
};
\addplot [black]
table {%
0 0
11 0
};
\path [draw=black, semithick]
(axis cs:0,0)
--(axis cs:0,1);

\addplot [semithick, black, mark=*, mark size=3, mark options={solid,fill=white}, only marks]
table {%
0 1
};
\addplot [semithick, black]
table {%
0 0
0 0
};
\end{axis}

\end{tikzpicture} & 
\begin{tikzpicture}

\definecolor{darkgray176}{RGB}{176,176,176}

\begin{axis}[
height=\figH,
hide x axis,
hide y axis,
tick align=outside,
tick pos=left,
width=\figW,
x grid style={darkgray176},
xmin=-1.1, xmax=12.1,
xtick style={color=black},
y grid style={darkgray176},
ymin=-1.1, ymax=1.1,
ytick style={color=black}
]
\path [draw=white, semithick]
(axis cs:0,0)
--(axis cs:0,-1);

\path [draw=white, semithick]
(axis cs:1,0)
--(axis cs:1,1);

\addplot [semithick, white, mark=*, mark size=3, mark options={solid}, only marks]
table {%
0 -1
1 1
};
\addplot [semithick, white]
table {%
0 0
1 0
};
\path [draw=black, semithick]
(axis cs:0,0)
--(axis cs:0,0.333333333333333);

\path [draw=black, semithick]
(axis cs:1,0)
--(axis cs:1,0);

\path [draw=black, semithick]
(axis cs:2,0)
--(axis cs:2,0);

\path [draw=black, semithick]
(axis cs:3,0)
--(axis cs:3,-0.333333333333333);

\path [draw=black, semithick]
(axis cs:4,0)
--(axis cs:4,0);

\path [draw=black, semithick]
(axis cs:5,0)
--(axis cs:5,0);

\path [draw=black, semithick]
(axis cs:6,0)
--(axis cs:6,0);

\path [draw=black, semithick]
(axis cs:7,0)
--(axis cs:7,0);

\path [draw=black, semithick]
(axis cs:8,0)
--(axis cs:8,0);

\path [draw=black, semithick]
(axis cs:9,0)
--(axis cs:9,0);

\path [draw=black, semithick]
(axis cs:10,0)
--(axis cs:10,0);

\path [draw=black, semithick]
(axis cs:11,0)
--(axis cs:11,0);

\addplot [semithick, black, mark=*, mark size=3, mark options={solid}, only marks]
table {%
0 0.333333333333333
1 0
2 0
3 -0.333333333333333
4 0
5 0
6 0
7 0
8 0
9 0
10 0
11 0
};
\addplot [black]
table {%
0 0
11 0
};
\path [draw=black, semithick]
(axis cs:0,0)
--(axis cs:0,0.333333333333333);

\addplot [semithick, black, mark=*, mark size=3, mark options={solid,fill=white}, only marks]
table {%
0 0.333333333333333
};
\addplot [semithick, black]
table {%
0 0
0 0
};
\end{axis}

\end{tikzpicture} \\
\begin{turn}{90} \hspace{0.05\textheight} $y[k]$ \end{turn}  & 
\begin{tikzpicture}

\definecolor{darkgray176}{RGB}{176,176,176}

\begin{axis}[
height=\figH,
hide x axis,
hide y axis,
tick align=outside,
tick pos=left,
width=\figW,
x grid style={darkgray176},
xmin=-1.1, xmax=12.1,
xtick style={color=black},
y grid style={darkgray176},
ymin=-1.1, ymax=1.1,
ytick style={color=black}
]
\path [draw=white, semithick]
(axis cs:0,0)
--(axis cs:0,-1);

\path [draw=white, semithick]
(axis cs:1,0)
--(axis cs:1,1);

\addplot [semithick, white, mark=*, mark size=3, mark options={solid}, only marks]
table {%
0 -1
1 1
};
\addplot [semithick, white]
table {%
0 0
1 0
};
\path [draw=black, semithick]
(axis cs:0,0)
--(axis cs:0,0);

\path [draw=black, semithick]
(axis cs:1,0)
--(axis cs:1,0);

\path [draw=black, semithick]
(axis cs:2,0)
--(axis cs:2,0);

\path [draw=black, semithick]
(axis cs:3,0)
--(axis cs:3,0.111111111111111);

\path [draw=black, semithick]
(axis cs:4,0)
--(axis cs:4,0.333333333333333);

\path [draw=black, semithick]
(axis cs:5,0)
--(axis cs:5,0.555555555555556);

\path [draw=black, semithick]
(axis cs:6,0)
--(axis cs:6,0.666666666666667);

\path [draw=black, semithick]
(axis cs:7,0)
--(axis cs:7,0.555555555555556);

\path [draw=black, semithick]
(axis cs:8,0)
--(axis cs:8,0.333333333333333);

\path [draw=black, semithick]
(axis cs:9,0)
--(axis cs:9,0.111111111111111);

\path [draw=black, semithick]
(axis cs:10,0)
--(axis cs:10,0);

\path [draw=black, semithick]
(axis cs:11,0)
--(axis cs:11,0);

\addplot [semithick, black, mark=*, mark size=3, mark options={solid}, only marks]
table {%
0 0
1 0
2 0
3 0.111111111111111
4 0.333333333333333
5 0.555555555555556
6 0.666666666666667
7 0.555555555555556
8 0.333333333333333
9 0.111111111111111
10 0
11 0
};
\addplot [black]
table {%
0 0
11 0
};
\path [draw=black, semithick]
(axis cs:0,0)
--(axis cs:0,0);

\addplot [semithick, black, mark=*, mark size=3, mark options={solid,fill=white}, only marks]
table {%
0 0
};
\addplot [semithick, black]
table {%
0 0
0 0
};
\end{axis}

\end{tikzpicture} & 
\begin{tikzpicture}

\definecolor{darkgray176}{RGB}{176,176,176}

\begin{axis}[
height=\figH,
hide x axis,
hide y axis,
tick align=outside,
tick pos=left,
width=\figW,
x grid style={darkgray176},
xmin=-1.1, xmax=12.1,
xtick style={color=black},
y grid style={darkgray176},
ymin=-1.1, ymax=1.1,
ytick style={color=black}
]
\path [draw=white, semithick]
(axis cs:0,0)
--(axis cs:0,-1);

\path [draw=white, semithick]
(axis cs:1,0)
--(axis cs:1,1);

\addplot [semithick, white, mark=*, mark size=3, mark options={solid}, only marks]
table {%
0 -1
1 1
};
\addplot [semithick, white]
table {%
0 0
1 0
};
\path [draw=black, semithick]
(axis cs:0,0)
--(axis cs:0,0);

\path [draw=black, semithick]
(axis cs:1,0)
--(axis cs:1,0);

\path [draw=black, semithick]
(axis cs:2,0)
--(axis cs:2,0);

\path [draw=black, semithick]
(axis cs:3,0)
--(axis cs:3,0.333333333333333);

\path [draw=black, semithick]
(axis cs:4,0)
--(axis cs:4,0.333333333333333);

\path [draw=black, semithick]
(axis cs:5,0)
--(axis cs:5,0);

\path [draw=black, semithick]
(axis cs:6,0)
--(axis cs:6,0);

\path [draw=black, semithick]
(axis cs:7,0)
--(axis cs:7,-0.333333333333333);

\path [draw=black, semithick]
(axis cs:8,0)
--(axis cs:8,-0.333333333333333);

\path [draw=black, semithick]
(axis cs:9,0)
--(axis cs:9,0);

\path [draw=black, semithick]
(axis cs:10,0)
--(axis cs:10,0);

\path [draw=black, semithick]
(axis cs:11,0)
--(axis cs:11,0);

\addplot [semithick, black, mark=*, mark size=3, mark options={solid}, only marks]
table {%
0 0
1 0
2 0
3 0.333333333333333
4 0.333333333333333
5 0
6 0
7 -0.333333333333333
8 -0.333333333333333
9 0
10 0
11 0
};
\addplot [black]
table {%
0 0
11 0
};
\path [draw=black, semithick]
(axis cs:0,0)
--(axis cs:0,0);

\addplot [semithick, black, mark=*, mark size=3, mark options={solid,fill=white}, only marks]
table {%
0 0
};
\addplot [semithick, black]
table {%
0 0
0 0
};
\end{axis}

\end{tikzpicture} & 
\begin{tikzpicture}

\definecolor{darkgray176}{RGB}{176,176,176}

\begin{axis}[
height=\figH,
hide x axis,
hide y axis,
tick align=outside,
tick pos=left,
width=\figW,
x grid style={darkgray176},
xmin=-1.1, xmax=12.1,
xtick style={color=black},
y grid style={darkgray176},
ymin=-1.1, ymax=1.1,
ytick style={color=black}
]
\path [draw=white, semithick]
(axis cs:0,0)
--(axis cs:0,-1);

\path [draw=white, semithick]
(axis cs:1,0)
--(axis cs:1,1);

\addplot [semithick, white, mark=*, mark size=3, mark options={solid}, only marks]
table {%
0 -1
1 1
};
\addplot [semithick, white]
table {%
0 0
1 0
};
\path [draw=black, semithick]
(axis cs:0,0)
--(axis cs:0,0);

\path [draw=black, semithick]
(axis cs:1,0)
--(axis cs:1,0);

\path [draw=black, semithick]
(axis cs:2,0)
--(axis cs:2,0);

\path [draw=black, semithick]
(axis cs:3,0)
--(axis cs:3,0.111111111111111);

\path [draw=black, semithick]
(axis cs:4,0)
--(axis cs:4,0.222222222222222);

\path [draw=black, semithick]
(axis cs:5,0)
--(axis cs:5,0.222222222222222);

\path [draw=black, semithick]
(axis cs:6,0)
--(axis cs:6,0.111111111111111);

\path [draw=black, semithick]
(axis cs:7,0)
--(axis cs:7,-0.111111111111111);

\path [draw=black, semithick]
(axis cs:8,0)
--(axis cs:8,-0.222222222222222);

\path [draw=black, semithick]
(axis cs:9,0)
--(axis cs:9,-0.222222222222222);

\path [draw=black, semithick]
(axis cs:10,0)
--(axis cs:10,-0.111111111111111);

\path [draw=black, semithick]
(axis cs:11,0)
--(axis cs:11,0);

\addplot [semithick, black, mark=*, mark size=3, mark options={solid}, only marks]
table {%
0 0
1 0
2 0
3 0.111111111111111
4 0.222222222222222
5 0.222222222222222
6 0.111111111111111
7 -0.111111111111111
8 -0.222222222222222
9 -0.222222222222222
10 -0.111111111111111
11 0
};
\addplot [black]
table {%
0 0
11 0
};
\path [draw=black, semithick]
(axis cs:0,0)
--(axis cs:0,0);

\addplot [semithick, black, mark=*, mark size=3, mark options={solid,fill=white}, only marks]
table {%
0 0
};
\addplot [semithick, black]
table {%
0 0
0 0
};
\end{axis}

\end{tikzpicture} 
\end{tabular}
\caption{An exemplary input sequence (top row, same for all three systems) is fed into an LTI system characterized by its impulse response (middle row) and leads to the shown output sequence (bottom row); the left column shows an averaging filter, an incoming impulse gets distributed on smaller constant values, the center column shows a backward difference, its impulse response has plus one for the slope upward from zero to one and minus one for the slope downward; the right column shows the series connection of both, averaging filter and backward difference;
unfilled circles denote signals at $t=0$.}
\label{fig_discrete_lti}
\end{figure}
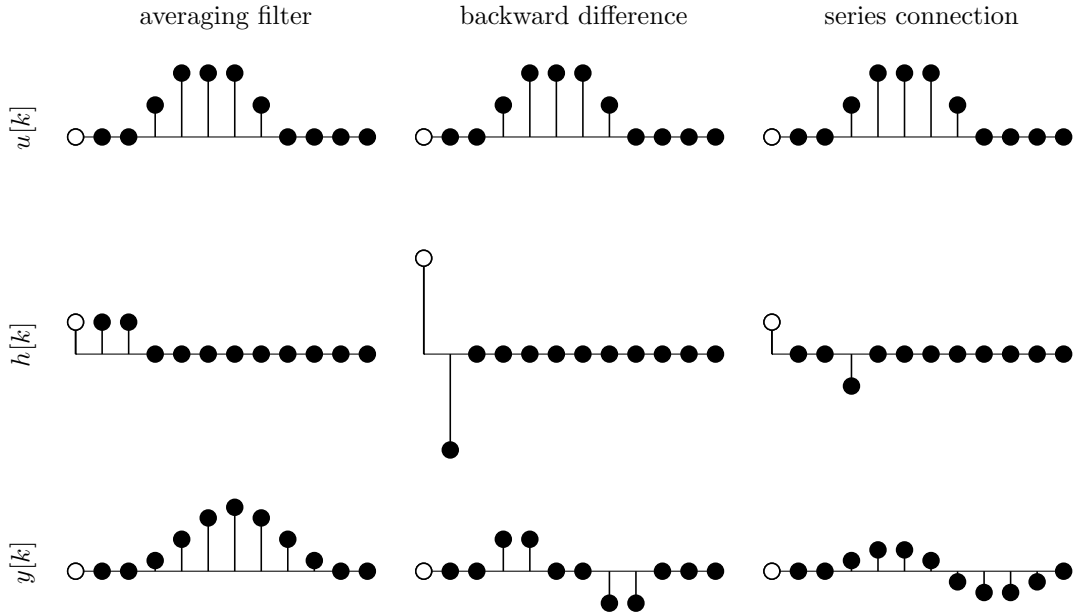
As \textsl{linear} implies, we may take advantage of the superposition principle, i.e. the system response to the sum  of two input signals is the sum of the separate responses.
Consequently, it makes sense to describe LTI-systems by their response to characteristic test-signals.
One of the most important test-signals is the unit impulse $\delta[k]$, a signal of unit strength at $k=0$ and zero elsewhere \cite{oppenheim2013discrete}.
If we have the system response $h[k]$ to this impulse, 
then we can compose the system response to any input signal, 
by representing the input signal as a sum of scaled and time-shifted unit impulses. 
Summing up each of the responses to the single input impulses $u[k]$ is a discrete convolution.
On physical grounds we emphasize the subset of causal LTI-systems, where causality means that an output is completely determined by current and previous states.
The sums for a general and specifically for a causal system, which starts at $k=0$ with vanishing initial conditions, read, respectively
\begin{subequations}
\label{eq_lti_dconv}
\begin{align}
y[n] &= \sum_{k=-\infty}^{\infty} u[k]\,h[n-k], \\
y[n] &= \sum_{k=0}^{n} u[k]\,h[n-k].
\end{align}
\end{subequations}
This justifies the name \textsl{transfer function} for the impulse response and at the same time it opens the way to a series connection of two systems (figure~\ref{fig_series_connection}).
As the response to a unit impulse of the first system, its transfer function, 
is the input to the second system, the output of the second system is the convolution of the impulse responses. 
In other words the transfer function of the series connection is the convolution of the transfer functions of its component systems.
For example we show the response of three systems, an averaging filter,
a backward difference and their series connection
\begin{align}
 y[k] &= \frac{u[k-2] + u[k-1] + u[k]}{3}&   &\leadsto&  h_\text{avg}[k]
 &= \left\{ \begin{array}{rl} 
 \frac{1}{3} & \text{for} \quad 0\le k \le 2 \\
 0 & \text{otherwise}
 \end{array}\right. \\
y[k]  &=  u[k] - u[k-1]&  &\leadsto& h_\text{bd}[k]&= \left\{ \begin{array}{rl} 
 1 & \text{for} \quad k=0  \\
-1 & \text{for} \quad k=1  \\
 0 & \text{otherwise}
 \end{array}\right. \\
 y[k] &= \frac{u[k] - u[k-3]}{3}& &\leadsto& h_\text{series}[k]&= \left\{ \begin{array}{rl} 
 \frac{1}{3} & \text{for} \quad k=0  \\[1mm]
-\frac{1}{3} & \text{for} \quad k=3  \\[1mm]
 0 & \text{otherwise}
 \end{array}\right. 
\end{align}
on an arbitrary input signal in figure~\ref{fig_discrete_lti}.
Note the similarity between adding impulse responses of LTI systems and the superposition of forces in the Boussinesq problem \eqref{eq:s_nonlocal}. 

\subsubsection{Continuous-time LTI-systems}
\begin{figure}
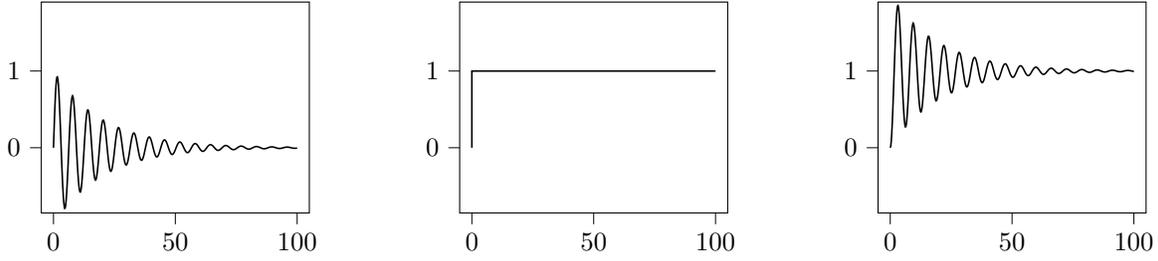

\setlength{\figW}{0.33\textwidth}
\setlength{\figH}{0.2\textheight}
\input{fig_pt2_response}
\hfill
\begin{tikzpicture}

\definecolor{darkgray176}{RGB}{176,176,176}

\begin{axis}[
height=\figH,
tick align=outside,
tick pos=left,
width=\figW,
x grid style={darkgray176},
xmin=-5, xmax=105,
xtick style={color=black},
y grid style={darkgray176},
ymin=-0.85, ymax=1.9,
ytick style={color=black}
]
\addplot [semithick, black]
table {%
0 0
0 1
0.200400801603206 1
0.400801603206413 1
0.601202404809619 1
0.801603206412826 1
1.00200400801603 1
1.20240480961924 1
1.40280561122244 1
1.60320641282565 1
1.80360721442886 1
2.00400801603206 1
2.20440881763527 1
2.40480961923848 1
2.60521042084168 1
2.80561122244489 1
3.0060120240481 1
3.2064128256513 1
3.40681362725451 1
3.60721442885772 1
3.80761523046092 1
4.00801603206413 1
4.20841683366733 1
4.40881763527054 1
4.60921843687375 1
4.80961923847695 1
5.01002004008016 1
5.21042084168337 1
5.41082164328657 1
5.61122244488978 1
5.81162324649299 1
6.01202404809619 1
6.2124248496994 1
6.41282565130261 1
6.61322645290581 1
6.81362725450902 1
7.01402805611222 1
7.21442885771543 1
7.41482965931864 1
7.61523046092184 1
7.81563126252505 1
8.01603206412826 1
8.21643286573146 1
8.41683366733467 1
8.61723446893788 1
8.81763527054108 1
9.01803607214429 1
9.21843687374749 1
9.4188376753507 1
9.61923847695391 1
9.81963927855711 1
10.0200400801603 1
10.2204408817635 1
10.4208416833667 1
10.6212424849699 1
10.8216432865731 1
11.0220440881764 1
11.2224448897796 1
11.4228456913828 1
11.623246492986 1
11.8236472945892 1
12.0240480961924 1
12.2244488977956 1
12.4248496993988 1
12.625250501002 1
12.8256513026052 1
13.0260521042084 1
13.2264529058116 1
13.4268537074148 1
13.627254509018 1
13.8276553106212 1
14.0280561122244 1
14.2284569138277 1
14.4288577154309 1
14.6292585170341 1
14.8296593186373 1
15.0300601202405 1
15.2304609218437 1
15.4308617234469 1
15.6312625250501 1
15.8316633266533 1
16.0320641282565 1
16.2324649298597 1
16.4328657314629 1
16.6332665330661 1
16.8336673346693 1
17.0340681362725 1
17.2344689378758 1
17.434869739479 1
17.6352705410822 1
17.8356713426854 1
18.0360721442886 1
18.2364729458918 1
18.436873747495 1
18.6372745490982 1
18.8376753507014 1
19.0380761523046 1
19.2384769539078 1
19.438877755511 1
19.6392785571142 1
19.8396793587174 1
20.0400801603206 1
20.2404809619238 1
20.4408817635271 1
20.6412825651303 1
20.8416833667335 1
21.0420841683367 1
21.2424849699399 1
21.4428857715431 1
21.6432865731463 1
21.8436873747495 1
22.0440881763527 1
22.2444889779559 1
22.4448897795591 1
22.6452905811623 1
22.8456913827655 1
23.0460921843687 1
23.2464929859719 1
23.4468937875751 1
23.6472945891784 1
23.8476953907816 1
24.0480961923848 1
24.248496993988 1
24.4488977955912 1
24.6492985971944 1
24.8496993987976 1
25.0501002004008 1
25.250501002004 1
25.4509018036072 1
25.6513026052104 1
25.8517034068136 1
26.0521042084168 1
26.25250501002 1
26.4529058116232 1
26.6533066132265 1
26.8537074148297 1
27.0541082164329 1
27.2545090180361 1
27.4549098196393 1
27.6553106212425 1
27.8557114228457 1
28.0561122244489 1
28.2565130260521 1
28.4569138276553 1
28.6573146292585 1
28.8577154308617 1
29.0581162324649 1
29.2585170340681 1
29.4589178356713 1
29.6593186372745 1
29.8597194388778 1
30.060120240481 1
30.2605210420842 1
30.4609218436874 1
30.6613226452906 1
30.8617234468938 1
31.062124248497 1
31.2625250501002 1
31.4629258517034 1
31.6633266533066 1
31.8637274549098 1
32.064128256513 1
32.2645290581162 1
32.4649298597194 1
32.6653306613226 1
32.8657314629259 1
33.0661322645291 1
33.2665330661323 1
33.4669338677355 1
33.6673346693387 1
33.8677354709419 1
34.0681362725451 1
34.2685370741483 1
34.4689378757515 1
34.6693386773547 1
34.8697394789579 1
35.0701402805611 1
35.2705410821643 1
35.4709418837675 1
35.6713426853707 1
35.8717434869739 1
36.0721442885772 1
36.2725450901804 1
36.4729458917836 1
36.6733466933868 1
36.87374749499 1
37.0741482965932 1
37.2745490981964 1
37.4749498997996 1
37.6753507014028 1
37.875751503006 1
38.0761523046092 1
38.2765531062124 1
38.4769539078156 1
38.6773547094188 1
38.877755511022 1
39.0781563126253 1
39.2785571142285 1
39.4789579158317 1
39.6793587174349 1
39.8797595190381 1
40.0801603206413 1
40.2805611222445 1
40.4809619238477 1
40.6813627254509 1
40.8817635270541 1
41.0821643286573 1
41.2825651302605 1
41.4829659318637 1
41.6833667334669 1
41.8837675350701 1
42.0841683366734 1
42.2845691382766 1
42.4849699398798 1
42.685370741483 1
42.8857715430862 1
43.0861723446894 1
43.2865731462926 1
43.4869739478958 1
43.687374749499 1
43.8877755511022 1
44.0881763527054 1
44.2885771543086 1
44.4889779559118 1
44.689378757515 1
44.8897795591182 1
45.0901803607214 1
45.2905811623246 1
45.4909819639279 1
45.6913827655311 1
45.8917835671343 1
46.0921843687375 1
46.2925851703407 1
46.4929859719439 1
46.6933867735471 1
46.8937875751503 1
47.0941883767535 1
47.2945891783567 1
47.4949899799599 1
47.6953907815631 1
47.8957915831663 1
48.0961923847695 1
48.2965931863727 1
48.496993987976 1
48.6973947895792 1
48.8977955911824 1
49.0981963927856 1
49.2985971943888 1
49.498997995992 1
49.6993987975952 1
49.8997995991984 1
50.1002004008016 1
50.3006012024048 1
50.501002004008 1
50.7014028056112 1
50.9018036072144 1
51.1022044088176 1
51.3026052104208 1
51.5030060120241 1
51.7034068136273 1
51.9038076152305 1
52.1042084168337 1
52.3046092184369 1
52.5050100200401 1
52.7054108216433 1
52.9058116232465 1
53.1062124248497 1
53.3066132264529 1
53.5070140280561 1
53.7074148296593 1
53.9078156312625 1
54.1082164328657 1
54.3086172344689 1
54.5090180360721 1
54.7094188376754 1
54.9098196392786 1
55.1102204408818 1
55.310621242485 1
55.5110220440882 1
55.7114228456914 1
55.9118236472946 1
56.1122244488978 1
56.312625250501 1
56.5130260521042 1
56.7134268537074 1
56.9138276553106 1
57.1142284569138 1
57.314629258517 1
57.5150300601202 1
57.7154308617234 1
57.9158316633267 1
58.1162324649299 1
58.3166332665331 1
58.5170340681363 1
58.7174348697395 1
58.9178356713427 1
59.1182364729459 1
59.3186372745491 1
59.5190380761523 1
59.7194388777555 1
59.9198396793587 1
60.1202404809619 1
60.3206412825651 1
60.5210420841683 1
60.7214428857715 1
60.9218436873748 1
61.122244488978 1
61.3226452905812 1
61.5230460921844 1
61.7234468937876 1
61.9238476953908 1
62.124248496994 1
62.3246492985972 1
62.5250501002004 1
62.7254509018036 1
62.9258517034068 1
63.12625250501 1
63.3266533066132 1
63.5270541082164 1
63.7274549098196 1
63.9278557114229 1
64.1282565130261 1
64.3286573146293 1
64.5290581162325 1
64.7294589178357 1
64.9298597194389 1
65.1302605210421 1
65.3306613226453 1
65.5310621242485 1
65.7314629258517 1
65.9318637274549 1
66.1322645290581 1
66.3326653306613 1
66.5330661322645 1
66.7334669338677 1
66.9338677354709 1
67.1342685370741 1
67.3346693386774 1
67.5350701402806 1
67.7354709418838 1
67.935871743487 1
68.1362725450902 1
68.3366733466934 1
68.5370741482966 1
68.7374749498998 1
68.937875751503 1
69.1382765531062 1
69.3386773547094 1
69.5390781563126 1
69.7394789579158 1
69.939879759519 1
70.1402805611223 1
70.3406813627254 1
70.5410821643287 1
70.7414829659319 1
70.9418837675351 1
71.1422845691383 1
71.3426853707415 1
71.5430861723447 1
71.7434869739479 1
71.9438877755511 1
72.1442885771543 1
72.3446893787575 1
72.5450901803607 1
72.7454909819639 1
72.9458917835671 1
73.1462925851703 1
73.3466933867736 1
73.5470941883768 1
73.74749498998 1
73.9478957915832 1
74.1482965931864 1
74.3486973947896 1
74.5490981963928 1
74.749498997996 1
74.9498997995992 1
75.1503006012024 1
75.3507014028056 1
75.5511022044088 1
75.751503006012 1
75.9519038076152 1
76.1523046092184 1
76.3527054108216 1
76.5531062124249 1
76.7535070140281 1
76.9539078156313 1
77.1543086172345 1
77.3547094188377 1
77.5551102204409 1
77.7555110220441 1
77.9559118236473 1
78.1563126252505 1
78.3567134268537 1
78.5571142284569 1
78.7575150300601 1
78.9579158316633 1
79.1583166332665 1
79.3587174348697 1
79.559118236473 1
79.7595190380761 1
79.9599198396794 1
80.1603206412826 1
80.3607214428858 1
80.561122244489 1
80.7615230460922 1
80.9619238476954 1
81.1623246492986 1
81.3627254509018 1
81.563126252505 1
81.7635270541082 1
81.9639278557114 1
82.1643286573146 1
82.3647294589178 1
82.565130260521 1
82.7655310621243 1
82.9659318637275 1
83.1663326653307 1
83.3667334669339 1
83.5671342685371 1
83.7675350701403 1
83.9679358717435 1
84.1683366733467 1
84.3687374749499 1
84.5691382765531 1
84.7695390781563 1
84.9699398797595 1
85.1703406813627 1
85.3707414829659 1
85.5711422845691 1
85.7715430861723 1
85.9719438877756 1
86.1723446893788 1
86.372745490982 1
86.5731462925852 1
86.7735470941884 1
86.9739478957916 1
87.1743486973948 1
87.374749498998 1
87.5751503006012 1
87.7755511022044 1
87.9759519038076 1
88.1763527054108 1
88.376753507014 1
88.5771543086172 1
88.7775551102204 1
88.9779559118237 1
89.1783567134269 1
89.3787575150301 1
89.5791583166333 1
89.7795591182365 1
89.9799599198397 1
90.1803607214429 1
90.3807615230461 1
90.5811623246493 1
90.7815631262525 1
90.9819639278557 1
91.1823647294589 1
91.3827655310621 1
91.5831663326653 1
91.7835671342685 1
91.9839679358717 1
92.184368737475 1
92.3847695390782 1
92.5851703406814 1
92.7855711422846 1
92.9859719438878 1
93.186372745491 1
93.3867735470942 1
93.5871743486974 1
93.7875751503006 1
93.9879759519038 1
94.188376753507 1
94.3887775551102 1
94.5891783567134 1
94.7895791583166 1
94.9899799599198 1
95.190380761523 1
95.3907815631263 1
95.5911823647295 1
95.7915831663327 1
95.9919839679359 1
96.1923847695391 1
96.3927855711423 1
96.5931863727455 1
96.7935871743487 1
96.9939879759519 1
97.1943887775551 1
97.3947895791583 1
97.5951903807615 1
97.7955911823647 1
97.9959919839679 1
98.1963927855711 1
98.3967935871744 1
98.5971943887776 1
98.7975951903808 1
98.997995991984 1
99.1983967935872 1
99.3987975951904 1
99.5991983967936 1
99.7995991983968 1
100 1
};
\end{axis}

\end{tikzpicture}
\hfill
\input{fig_pt2_integrator_response}
\caption{Impulse response of a PT2 system (damped harmonic oscillator), an integrator and the series connection of both (from left to right).}
\label{fig_continuous_lti}
\end{figure}
The continuous-time unit impulse, the Dirac delta distribution, has the same meaning as its discrete-time counterpart.
It characterizes a LTI system and we can use it to represent an arbitrary signal as impulse train and thus find its response \cite{dorf2017modern}.
In contrast to equation~\eqref{eq_lti_dconv} superposition means now finite summation turns into integration. 
The convolution inegrals for general and specifically for causal systems with vanishing initial conditions from $t=0$ on read, respectively
\begin{subequations}
\label{eq_lti_cconv}
\begin{align}
 y(t) = & \int_{-\infty}^{\infty} u(\tau) h(t-\tau) \,\text{d}\tau ,\\
 y(t) = & \int_{0}^t u(\tau) h(t-\tau) \,\text{d}\tau.
 \label{eq_lti_cconv_causal}
\end{align}
\end{subequations}
The physical correspondence for causality in space means a preferential direction along which effects may depend on each other, e.g. the right point affects the left point but not vice versa, which is uncommon.
Also memoryless in time refers to a direction, from past to future, in which dependencies may exist.
Often, tacitly assuming causality, memoryless means, each input affects just the current output, which corresponds to \textsl{local} in space, when each force affects only the immediate vicinity.  

For a series connection we may again subject the first system to the unit impulse and feed its output into the second system, revealing that the impulse response of the compound is the convolution of the separate impulse responses. 
For illustration, figure~\ref{fig_continuous_lti} shows the impulse responses of a PT2 system, in mechanics known as harmonic oscillator, an integrator 
\begin{align}
    \ddot{y} + 2 \delta \dot{y} + \omega_0^2 y &= u(t) 
    \qquad\text{with}\quad \delta=0.05 \qquad\text{and}\quad \omega_0=1.0,\\
    \dot{y} &= u(t),
\end{align}
and the series connection of both of them, all with vanishing initial conditions.

In the common description with Laplace-transform, you will notice that the Laplace-transform of the unit impulse is unity and thus the impulse response is the transfer function as it was for discrete-time systems.
Further you can show by the definition of the Laplace transform, that convolution in the time domain corresponds to multiplication in the Laplace domain.
This finding literally raises the power of Laplace transform tables, 
as you do not need to tabulate every specific expression as entry in the table, but can find it as product of entries, beyond that it may generally help to solve and understand integrals over products\footnote{\href{https://www.youtube.com/watch?v=851U557j6HE}{3Blue1Brown: Researchers thought this was a bug (Borwein integrals)}}.
By the way, many transfer functions are rational functions and on a multiplication of them, as for a series connection, you again encounter convolution.
There are polynomials in numerator and denominator of the transfer functions and multiplication of them corresponds to a discrete convolution of their coefficients as mentioned in section~\ref{sec_defprop}.
Once you got to know something, you will find it everywhere.

\subsection{Viscoelasticity}
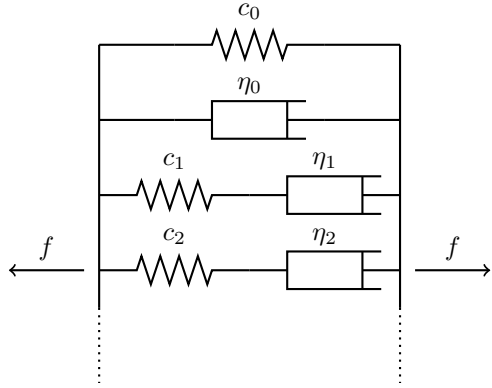
\begin{figure}
\centering
\begin{tikzpicture}
\draw[thick] (0, 3) -- (1, 3) (3, 3) -- (4, 3); 
\draw (1, 3.0) pic [scale=0.5, thick] {DKspring=4};
\draw (2, 3.45) node {$c_0$};
\draw[thick] (0, 2) -- (1, 2) (3, 2) -- (4, 2); 
\draw (1, 2.0) pic [scale=0.5, thick] {DKdashpot=4}; 
\draw (2, 2.45) node {$\eta_0$};
\draw (0, 1.0) pic [scale=0.5, thick] {DKspring=4};
\draw (2, 1.0) pic [scale=0.5, thick] {DKdashpot=4};
\draw (1, 1.45) node {$c_1$};
\draw (3, 1.45) node {$\eta_1$};
\draw (0, 0.0) pic [scale=0.5, thick] {DKspring=4};
\draw (2, 0.0) pic [scale=0.5, thick] {DKdashpot=4}; 
\draw (1, 0.45) node {$c_2$};
\draw (3, 0.45) node {$\eta_2$};
\draw[thick] (0, 3) -- (0, -0.5);
\draw[thick] (4, 3) -- (4, -0.5);
\draw[thick, ->] (4.2, 0) -- node[above]{$f$} (5.2, 0);
\draw[thick, ->] (-0.2, 0) -- node[above]{$f$} (-1.2, 0);
\draw[dotted, thick] (0,-0.5) -- (0,-1.5);
\draw[dotted, thick] (4,-0.5) -- (4,-1.5);
\end{tikzpicture}
 \caption{Rheological model of a viscoelastic material.}
 \label{fig_rheological_viscoelasticity}
\end{figure}
In a viscoelastic material, we observe both elastic and viscous effects.
Elasticity is represented by springs and its force depends on displacement, 
whereas viscosity is represented by dashpots and depends on the velocity.
In the linear case both forces read, respectively
\begin{align}
\label{eq_spring}  
f_\text{e} &= c u_\text{e}, \\                                                                                           
\label{eq_dashpot}  
f_\text{v} &= \eta \dot{u}_\text{v},                                                                                            \end{align}
where $c$ denotes spring stiffness and $\eta$ damping coefficient.
Rheological models describe materials as a compound of springs in series and in parallel, in plasticity there may be frictional elements too \cite{haupt2013continuum}.
It is customary to describe complex viscoelastic materials by a parallel connection of a spring (purely elastic), a dashpot (purely viscous) and further branches with spring and dashpot in series, 
as illustrated in figure~\ref{fig_rheological_viscoelasticity}.
Such a series connection of a spring and a dashpot is called Maxwell element and characterized by its relaxation time
\begin{equation}
 t_\text{rx} = \frac{\eta}{c},
\end{equation}
which describes how fast its stress decays after a jump in its length (relaxation test). 
Note that spring and dashpot are limit cases with a relaxation time of zero and infinity, respectively.

We are interested in the relation between total force $f(t)$ and displacement $u(t)$ of the parallel connection in a complex viscoelastic material (figure~\ref{fig_rheological_viscoelasticity}).
However, we must account for internal variables, the divisions into elastic and viscous displacement in each Maxwell element, which depend on the history.

Let's first analyze the force response of a single Maxwell element with $c,\eta > 0$, from which the model is composed.
There are two variables, stretch of spring $u_\text{e}$ and stretch of dashpot $u_\text{v}$.
They are related by total stretch and force consistency
\begin{align}
 u &= u_\text{e} + u_\text{v},  \\ 
 f &= c u_\text{e} = \eta \dot{u} _\text{v}. 
 \label{eq_force_consistent}
\end{align}
Since we are looking for a relation between total stretch and force, we eliminate $u_\text{e}$ and obtain a linear, nonhomogenous, ordinary differential equation (ODE)
\begin{equation}
 \dot{u}_\text{v} + \beta u_\text{v} = \beta u,
\end{equation}
with $\beta=c/\eta=t_\text{rx}^{-1}$.
Its solution of the homogenous equation and total solution, e.g. by variation of constants (actually for a first-order ODE there is only one constant), respectively, read
\begin{align}
 u_\text{v}^\text{h} &= C e^{-\beta t} \quad \text{with} \quad C=\text{const.}, \\
 \label{eq_uv_total}
 u_\text{v} &= u_\text{v}(t_0) e^{-\beta(t-t_0)}
 + \int_{t_0}^{t} u(\tau) \beta e^{-\beta(t-\tau)}\,\text{d}\tau,
\end{align}
where the total solution satisfies the initial condition $u_\text{v}(t_0)$.
Here the convolution integral emerges from the integration of the first order ODE.
Comparing with continuous LTI-systems \eqref{eq_lti_cconv_causal} we may note, that the solution of the homogenous equation with initial conditions $u_\text{v}^\text{h}(t_0)=\beta$ corresponds to the impulse response.
Anyway, here we found it more intuitive to approach the solution from calculus. 
Plugging solution~\eqref{eq_uv_total} into equation~\eqref{eq_force_consistent} we obtain the force
\begin{equation}
\label{eq_fu1}
 f(t) = c\Bigl( u(t) - u_\text{v}(t_0) e^{-\beta(t-t_0)}
 - \int_{t_0}^{t} u(\tau) \beta e^{-\beta(t-\tau)}\,\text{d}\tau \Bigr).
\end{equation}
Basically we were done, but in view of a parallel connection, as in figure~\ref{fig_rheological_viscoelasticity}, we would like to arrange \eqref{eq_fu1}
in a more convenient way. Our goal is a separation into initial value terms, to be evalutated once, and a collection of the remaining terms in one integral, instead of an integral for each Maxwell element.
Therefore we take advantage of the relation
\begin{equation}
\label{eq_u2exp}
 \int_{t_0}^t u(t) \beta e^{-\beta(t-\tau)}\,\text{d}\tau 
 = u(t)\big|e^{-\beta(t-\tau)}\big|_{\tau=t_0}^{t} = u(t) - u(t) e^{-\beta(t-t_0)},
\end{equation}
where we note that $u(t)$ is just a factor, since $t$ is not the integration variable.
Evaluating $u(t)$ from \eqref{eq_u2exp} in~\eqref{eq_fu1} we obtain
\begin{equation}
 f(t) = c  \bigl( u(t) - u_\text{v}(t_0)\bigr) e^{-\beta(t-t_0)}
 +  \int_{t_0}^{t} c \bigl( u(t) - u(\tau)  \bigr)  \beta e^{-\beta(t-\tau)}\,\text{d}\tau . 
\end{equation}
Finally, the trivial identity
\begin{equation}
\label{eq_u0_uv0}
 u(t) = u(t) + u(t_0) - u(t_0)
\end{equation}
allows to identify the same structure for the response of a complex material as for a single Maxwell element in terms of $u(t)$.
For the parallel connection of a spring (stiffness $c_0$), a dashpot (damping coefficient $\eta_0$)
and $N$ Maxwell elements we obtain
\begin{multline}
f(t) = c_0 u + \eta_0 \dot{u} 
+ \sum_{n=1}^N c_n\Bigl( u(t_0) - u_{\text{v}n}(t_0)\Bigr) e^{-\beta_n(t-t_0)} \\
+ K(t-t_0) \bigl(u(t)-u(t_0)\bigr) - \int_{t_0}^{t} K'(t-\tau)\bigr( u(t) - u(\tau)\bigr) \,\text{d}\tau, 
\end{multline}
with $K(x)=\sum\limits_{n=1}^N c_n e^{-\beta_n x}$ and its derivative $K'(x)=-\sum\limits_{n=1}^N c_n\beta_n e^{-\beta_n x}$.
Note that the Prony series $K(x)$ corresponds to a relaxation function and the kernel $K'(x)$ includes the history.
Further $K(x)$ may be seen as a finite approximation of a continuous spectrum \cite{krawietz2013materialtheorie}.

\subsection{Further applications}
Convolutions can be generalized to higher dimensions.
This is apparent from section \ref{sec_point_halfspace} about forces on a half-space.
Two-dimensional applications occur in image processing e.g. for deblurring.
Similarly, they are applied (in combination with pooling) in convolutional neural-networks, to extract information, e.g. edges, from images and thus reduce the problem complexity.

Another type of analysis is deconvolution, commonly referred to as \textsl{inverse problem}. 
Knowing one input function and the output of the convolution, the problem is to find a good (non-unique) approximation for the second input function. 
A practical example from seismology is the approximation of the position-dependent reflectivity from a seismogram and the wave signal, stipulating that the
measurement (seismogram) is the convolution of wave signal and Earth's reflectivity \cite{robinson1957predictive}.


\end{document}